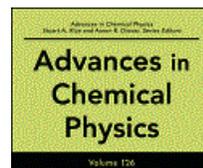

# Diamondoid Molecules

By:

### G.Ali Mansoori

*Departments of BioEngineering & Chemical Engineering*

*University of Illinois at Chicago*

(m/c 063) Chicago, IL 60607-7052

## Abstract

Diamondoids have been of great interest in recent years due to their role in nanotechnology, drug-delivery and medicine. In this review paper we introduce at first the cage nature of diamondoid molecules (polymantanes, adamantologues), the variety of their crystalline lattice structures, the nature of their structural isomers, their stereoisomers, and their other molecular specificities are presented. The carbon-carbon framework of diamondoids constitutes the fundamental repeating unit in the diamond lattice structure. It is demonstrated that diamondoids are very stable compounds.

The unique physicochemical properties of diamondoids due to their exceptional atomic arrangements including their melting points, molar enthalpies, molar entropies, molar heat capacities, vapor pressures and other phase transitions and solubilities data are reported and analyzed.

The Lewis acid–catalyzed rearrangement of hydrocarbons to synthesize lower diamondoids is presented and its limitations for the synthesis of higher diamondoids are discussed.

The natural occurrence of diamondoids in petroleum fluids and how they come to be present in such fluids is introduced. Field experiences of phase transitions and depositions as well as techniques for separation, detection and measurement of diamondoids from petroleum fluids is presented and discussed..

It is demonstrated that due to their six or more linking groups diamondoids have found major applications as templates and as molecular building blocks in polymers synthesis, nanotechnology, drug delivery, drug targeting, DNA directed assembly, DNA-amino acid nanostructure formation and in host-guest chemistry.

_____________________________________

(*) E-Mail: *mansoori@uic.edu*.

## Molecular Structure of Diamondoids

Diamondoid molecules are cage-like, ultra-stable, saturated hydrocarbons. These molecules are ringed compounds, which have a diamond-like structure consisting of a number of six-member carbon rings fused together (see Figure 1). More explicitly, they consist of repeating units of ten carbon atoms forming a tetra-cyclic cage system [1-3]. They are called "diamondoid" because their carbon-carbon framework constitutes the fundamental repeating unit in the diamond lattice structure. This structure was first determined in 1913 by Bragg and Bragg [4] using X-ray diffraction analysis.

The first and simplest member of the diamondoids group, adamantane, is a tricyclic saturated hydrocarbon (tricyclo[3.3.1.1]decane). Adamantane is followed by its polymantane homologues (adamantologues): diamantane, tria-, tetra-, penta- and hexamantane. Figure 1, illustrate the smaller diamondoid molecules, with the general chemical formula $C_{4n+6}H_{4n+12}$: adamantane ($C_{10}H_{16}$), diamantane ($C_{14}H_{20}$), and triamantane ($C_{18}H_{24}$),.. These lower adamantologues, each has only one isomer.

Depending on the spatial arrangement of the adamantane units, higher polymantanes (n≥4) can have several isomers and non-isomeric equivalents. There are three possible tetramantanes all of which are isomeric. They are depicted in Figure 2 as iso-, anti- and skew-tetramantane respectively. Anti- and skew-tetramantanes each possesses two quaternary carbon atoms, whereas iso-tetramantane has three.

With regards to the remaining members of the diamondoids group, there are seven possible pentamantanes, six of which are isomeric ($C_{26}H_{32}$) and obeying the molecular formula of the homologous series and one is non-isomeric ($C_{25}H_{30}$). For hexamantane, there are 24 possible structures: 17 are regular cata-condensed isomers with the chemical formula ($C_{30}H_{36}$), six are irregular cata-condensed isomers with the chemical formula ($C_{29}H_{34}$), and one is peri-condensed with the chemical formula ($C_{26}H_{30}$) as it is shown in Figure 3. The top and side views of a peri-condensed cyclohexamantane cage hydrocarbon are illustrated in Table 1.  The table also lists some physical properties of diamondoids mostly compiled by Chevron Texaco.

Diamondoids, when in solid state, diamondoids melt at much higher temperatures than other hydrocarbon molecules with the same number of carbon atoms in their structure. Since they also possess low strain energy they are more stable and stiff resembling diamond in a broad sense.  They contain dense, three- dimensional networks of covalent bonds, formed chiefly from first and second row atoms with a valence of three or more. Many of the diamondoids possess structures rich in tetrahedrally-coordinated carbon. They are materials with superior strength-to-weight ratio.

It has been found that adamantane crystallizes in a face-centered cubic lattice, which is extremely unusual for an organic compound. The molecule therefore should be completely free from both angle strain (since all carbon atoms are perfectly tetrahedral) and torsional strain (since all C-C bonds are perfectly staggered), making it a very stable compound and an excellent candidate for various applications as it will be discussed later in this report.

At the initial of growth, crystals of adamantane show only cubic and octahedral faces. The effects of this unusual structure on physical properties are interesting [5].

Many of the diamondoids can be brought to macroscopic crystalline forms with some special properties. For example, in it's crystalline lattice, the pyramidal-shape [1(2,3)4]pentamantane (see Table I) has a large void in comparison to similar crystals. Although it has a diamond-like macroscopic structure, it possesses the weak, non-covalent, intermolecular van der Waals attractive forces involved in forming a crystalline lattice [6,7]. Another example is the crystalline structure of 1,3,5,7-tetracarboxy adamantine, which is formed via carboxyl hydrogen bonds of each molecule with four tetrahedral nearest-neighbors. The similar structure in 1,3,5,7-tetraiodoadamantane crystal would be formed by I...I interactions. In 1,3,5,7-tetrahydroxyadamantane, the hydrogen bonds of hydroxyl groups produce a crystalline structure similar to inorganic compounds, like cesium chloride (*CsCl*) lattice [8] (see Figure 4).

The presence of chirality is another important feature in many derivatives of diamondoids. It should be pointed out that tetramantane[123] is the smallest lower diamondoids to possess chirality [6] (see Table I for chemical structures).

The vast number of structural isomers and stereoisomers is another property of diamondoids. For instance, octamantane possesses hundreds of isomers in five molecular weight classes. The octamantane class with formula $C_{34}H_{38}$ and molecular weight 446 has 18 chiral and achiral isomeric structures. Furthermore, there is unique and great geometric diversity with these isomers. For example rod-shaped diamondoids (with the shortest one being 1.0 nm long), disc-shaped and screw-shaped diamondoids (with different helical pitches and diameters) have been recognized [6], for example, as shown in Table I.

## Chemical and Physical Properties of Diamondoids

Diamondoids show unique properties due to their exceptional atomic arrangements. Adamantane consists of cyclohexane rings in 'chair' conformation. The name adamantane is derived from the Greek language word for diamond since its chemical structure is like the three-dimensional diamond subunit as shown in Figure 5.

Later the name diamondoids was chosen for all the higher cage hydrocarbon compounds of this series because they have the same structure as the diamond lattice: highly symmetrical and strain free so that their carbon atom structure can be superimposed upon a diamond lattice, as shown in Figure 5 for adamantane, diamantane and trimantane. These compounds are also known as adamantologues and polymantanes.

These compounds are chemically and thermally stable and strain-free. These characteristics cause high melting points (m.p.) in comparison to other hydrocarbons. For instance, the m.p. of adamantane is estimated to be ~269 °C, yet it sublimes easily, even at atmospheric pressure and room temperature. The melting point of diamantane is about 236.5 °C and the melting point of triamantane is estimated to be 221.5 °C. The available melting point data for diamondoids are reported in Table I.

Limited amounts of other chemical and physical property data have been reported in the literature for diamondoids [5, 9-30]. What is available is mostly for low molecular weight diamondoids. In what follows we report and analyze a selection of the available property data for diamondoids.

Adamantane {(CAS No: 281-23-2) 1-tricyclo[3.3.1.1$^{3,7}$]decane} is a cage hydrocarbon with a white or almost white crystalline solid nature, like solid wax, at normal conditions. Its odor resembles that of camphor. It is a stable and non-biodegradable compound, that is combustible due its hydrocarbon nature. It has not been found to be hazardous or toxic to living entities [14,15]. It should be pointed out that adamantane can exist in gas, liquid and two solid crystalline states.

Diamantane {(CAS No: 2292-79-7) pentacyclo[7.3.1.1$^{4.12}$.0$^{2.7}$.0$^{6.11}$]tetradecane} also known as {decahydro-3,5,1,7-[1.2.3.4]-butanetetraylnaphtalene } can exist in gas, liquid and three different solid crystalline states. Higher diamondoids possess two or more solid crystalline states [5].

The solid and liquid vapor pressures for adamantane and diamantane have been determined between ambient temperature and their estimated critical points using various measurement techniques by a number of investigators [13,16,17]. In Table II, equations representing the natural logarithm of adamantane and diamantane vapor pressures, *ln* P, as a function of absolute temperature as measured by various investigators are reported along with the temperature ranges of their validity. Figures (6) and (7) are based on the vapor pressure data reported in Table II. In these figures the smoothed vapor-liquid-solid phase transition lines (vapor pressures) for adamantane and diamantane are reported. There is a limited amount of data for vapor pressures of binary mixtures of adamantane and diamantane with other hydrocarbons available [18].

In Table III we also report the average of the available data for molar enthalpies, molar entropies and molar heat capacities of adamantane and diamantane as measured and reported by various investigators [10,19 - 30].

In Table IV we report the available enthalpies of formation, sublimation and combustion of methyl-adamantanes, dimethyl-adamantane, trimethyl-adamantane and tetramethyl-adamantane and compare them with the same properties of adamantane as measured by various investigators [29,30]. Also reported in Table IV are the same properties for 1-methyl-diamanten, 3-methyl-diamantane and 4-methyl-diamantane as compared with the diamantane data.

A number of other thermodynamic properties of adamantane and diamantane in different phases are reported by Kabo et al [5]. They include: (i). Standard molar thermodynamic functions for adamantane in the ideal gas state as calculated by statistical thermodynamics methods; and (ii). Temperature dependence of the heat capacities of adamantane in the condensed state between 340 and 600 K as measured by a scanning calorimeter and reported here in Figure 8. According to this figure liquid adamantane converts to a solid plastic with simple cubic crystal structure upon freezing. After further cooling it moves into another solid state, an *fcc* crystalline phase.

For higher diamondoids very limited data are available in the literature. For triamantane, for example, the enthalpy and entropy of transition from crystalline phase II to crystalline phase I are reported [31] as follows:

$\Delta H_{phase\ II–I}$      1.06     kJ/mol  (at 293.65°K)
$\Delta S_{phase\ II–I}$      3.77     kJ/mol  (at 293.65°K)

Little data is reported on tetra-, penta- and hexamantane and other higher diamondoids. What is available is compiled by ChevronTexaco scientists as reported in Table I. This is possibly due to the fact that of these compounds only anti-tetramantane has been successfully synthesized in the laboratory in small quantities [32,33].

**Solubilities:** A limited amount of solubility data for diamondoids in liquid solvents is available and they are reported in Table V. In this table the solubility limits of adamantane and diamantane in various liquid solvents at normal conditions are reported. In producing this data a known amount of diamondoid was titrated with various liquids at 25°C. Continuous stirring was used until all the diamondoid was dissolved, which defined the solubility limit [13]. According to Table V adamantane solubility in tetrahydrofuran (THF) is higher than in other organic solvents. Overall cyclohexane is a better solvent for diamondoids among the liquids tested due to the similarities in the molecular structure of cyclohexane to diamondoids. It should be also pointed out that since diamondoid molecules are substantially hydrophobic their solubility in organic solvents is a function of their hydrophobicity [13]. Carbon atoms on diamondoid surfaces possess primary carbons in the methylated analogs, as well as secondary and tertiary carbons. This makes it possible to use a wide and selective array of derivatization strategies on diamondoids [34]. Diamondoids' derivatization could enhance their solubility and as a result their potential applications in separation schemes, chromatography and pharmaceuticals.

The solubilities of adamantane and diamantane in supercritical (dense) methane, ethane and carbon dioxide gases have been measured by a number of investigators [35-37] at a few temperatures with various pressures and solvent densities. These measurements are reported by Figures (9)-(12).

Experimental data [36] on the effect of temperature and pressure on the supercritical solubility of adamantane in dense (supercritical) carbon dioxide gas is reported by Figure (9).

In Figure (10) the experimental data [35,37] showing the effect of temperature and supercritical solvent density on the solubility of adamantane in dense (supercritical) carbon dioxide at various temperatures are reported.

The data of Ref. [35] is reported graphically by Figure 11 and shows the effect of pressure on the solubility of adamantane in various supercritical solvents (carbon dioxide, methane and ethane) at 333 K.

A graphical representation of diamantane solubility data ~~of~~ [36] in various supercritical solvents (carbon dioxide and ethane at 333 K and methane at 353 K) is shown in Figure 12.

Trends of solubility enhancement for each diamondoid follow regular behavior like other heavy hydrocarbon solutes in supercritical solvents with respect to variations in pressure and density [38,39]. Supercritical solubilities of these lower diamondoids have been successfully correlated through cubic equations of state [35].

The supercritical fluid and liquid solubilities reported in Figures 9-12 suggest that diamondoids will preferentially partition themselves into the high pressure, high temperature and rather low-boiling fraction of any mixture including crude oil.

## Synthesis of Diamondoids

Originally diamondoids were considered as hypothetical molecules since they could neither be isolated from a natural repository nor made through rational organic synthesis [3] until 1933 when adamantane was discovered and isolated from petroleum [40]. The natural source of diamondoids was the only effective source until 1941 when adamantane was synthesized for the first time, though the yield was very low [32,41,42]. In 1957 Schleyer introduced a Lewis acid–catalyzed rearrangement of hydrocarbons to produce adamantane [43,44]. A Lewis acid is any acid that can accept a pair of electrons and form a coordinate covalent bond [45]. According to Schleyer *endo*-trimethylenenorbornane, which could be produced readily, rearranged to adamantane when refluxed overnight with aluminum bromide or aluminum chloride as shown in Figure 13. Later, in 1965, Cupas and Schleyer [46] used the same Lewis acid–catalyzed rearrangement synthesis approach to produce diamantane.

Schleyer's Lewis acid–catalyzed rearrangement method, which is based on diamondoid thermodynamic stability during carbocation rearrangements, has had little or no success to synthesize diamondoids beyond triamantane. In recent years, outstanding successes have been achieved in the synthesis of adamantane and other lower molecular weight diamondoids [42,49]. Some new methods have been developed and the yield has been increased to 60%.

While attempts to synthesize lower diamondoids (adamantane, diamantane, triamantane) have been successful through Lewis acid-catalyzed rearrangement, it is no longer a method of choice for synthesis of higher polymantanes [32]. The forth member of diamondoid homologues, tetramantane ($C_{22}H_{28}$), has three isomers. The fifth member, pentamantane ($C_{26}H_{32}$), has six isomers while the sixth member, hexamantane ($C_{30}H_{36}$), has as many as seventeen possible isomers. Producing these heavier diamondoids via the above-mentioned rearrangement synthesis method has not been convenient because of their close structural properties and their lower thermal stability.

The synthesis of higher diamondoids, in its nature, is a challenging and complex process. For example, after numerous efforts the anti isomer of tetramantane was synthesized with only 10 % yield [32]. The usage of zeolites as the catalyst in synthesis of adamantane has been investigated and different types of zeolites have been tested for achieving better catalyst activity and selectivity in adamantane formation reactions [42]. Recently, Shibuya et al. have represented two convenient methods for synthesis of enantiomeric adamantane derivatives [50]. High molecular weight diamondoids have not been synthesized as of this date. Present day efforts are towards more economical separation of high molecular weight diamondoids from petroleum fluids and more economical synthesis of lower molecular weight diamondoids [51].

## Natural Occurrence of Diamondoids

Adamantane and other diamondoids are constituents of petroleum, gas condensate (also called NGL or natural gas liquid) and natural gas reservoirs [52-56]. Adamantane was originally discovered [40] and isolated from petroleum fractions of the Hodonin oilfields in Czechoslovakia in 1933.

Naturally occurring adamantane is generally accompanied by small amounts of alkylated adamantane: 2-methyl-; 1-ethyl-; and probably 1-methyl-; 1,3-dimethyl- adamantane; and others [3]. Diamantane, triamantane and their alkyl-substituted compounds are also present in certain petroleum crude oils. Their concentrations in crude oils are generally lower than that of adamantane and its alkyl-substituted compounds.
Tetramantane, Pentamantanes and hexamantanes are found in some deep natural gas deposits [57,58] but are not readily synthesized in the laboratory.

The question of how diamondoids (polymantanes) come to be present in petroleum fluids is an interesting one. Diamond in nature is formed from abiogenic carbon and is abiogenic. However, the structurally related diamondoids in oil are biogenic. Diamondoids in petroleum are believed to be formed from enzymatically-created lipids with subsequent structural rearrangement during the process of source rock maturation and oil generation. Because of this, the diamondoid content of petroleum is applied to distinguish source rock facies [59-62].

Due to particular structure of diamondoids, they could be useful in making new biomarkers with more stability than the existing ones. New findings indicate that

_______________________

diamondoids are the appropriate alternatives for analyzing reservoirs which could not be assessed with conventional techniques. They appear to be resistant to biodegradation. Following biodegradation the remaining oil is enriched with diamondoids. Then the level of biodegradation will be estimated by determination of the ratio of diamondoids to their derivatives, particularly when main part of hydrocarbons has been degraded [556,59-62].

It is believed that [57,63] the diamondoids found in petroleum result from carbonium ion rearrangements of suitable organic precursors (such as multi-ringed terpene hydrocarbons) on clay mineral from the same source. In view of the Lewis acid-catalyzed isomerization (rearrangement) of hydrocarbons as discussed above, it is speculated that diamondoids may have been formed via homologation of the lower adamantologues at high pressure and temperature in the natural underground oil and gas reservoirs. The lower adamantologues are believed to have been formed originally by the catalytic rearrangement of tricycloalkanes (Figure 13) during or after oil generation [64].

Despite the above-mentioned hypothesis aimed at finding the adamantane precursors, the adamantane-forming substance in natural petroleum fluids remains unknown. In rare cases, tetra-, penta-, and hexamantanes are also found in petroleum crude oils. Tetramantane, pentamantane and hexamantane, were discovered for the first time in 1995 in a gas condensate (NGL) produced from a very deep (~ 6800 m below the surface) petroleum reservoir located in the US Gulf Coast [64]. A group of investigators [65] recently reported isolation of crystals of the lower-order diamondoid cyclohexamantane ($C_{26}H_{30}$), Figure (3), from distilled Gulf of Mexico petroleum using reverse-phase HPLC (high-performance liquid chromatography). They determined the structure of $C_{26}H_{30}$ using X-ray diffraction, mass spectroscopy, and 1H,13C NMR spectroscopy. They also used the experimental Raman spectra of crystalline diamond, adamantane, and nanophase diamond to indirectly identify frequencies in the experimental Raman spectra of $C_{26}H_{30}$.

Recent identification and isolation of crystals of many new medium and higher-order diamondoids from petroleum have been reported [6]. These investigators reported the isolation of multiple families of higher diamondoid molecules containing 4 to 11 diamond-crystal cages from petroleum and they supplied X-ray structures for representatives from three families. They reported separation of higher diamondoids from selected petroleum feedstock by distillation. They also reported the removal of non-diamondoids by pyrolysis at 400 – 450 ºC. Aromatic and polar compounds were removed from pyrolysis products by argentic silica gel liquid chromatography. Higher diamondoids were then isolated by a combination of reversed-phase HPLC on octadecyl silane columns and highly shape-selective Hypercarb® high-performance liquid chromatography (HPLC) columns. Individual higher diamondoids were re-crystallized to high purity [6].

A group of investigators recently suggested that the density-functional theory (DFT) calculated IR and Raman spectra is a useful tool for direct characterization of the structures of diamondoids with increasing complexity [66]. They applied DFT to calculate Raman spectra whose frequencies and relative intensities were shown to be in excellent agreement with the experimental Raman spectra for $C_{26}H_{30}$, thus providing direct vibrational proof of its existence.

Characteristics of diamondoids present in petroleum fluids are considered a proof of the biogenic origin of petroleum. Since diamondoids are resistant to biodegradation, diamondoid compounds and their application for the characterization of heavily biodegraded oils has been suggested and demonstrated. For example, it is found that the kinds of diamondoids present in crude oil are indicators of the level of petroleum biodegradation, which are not affected by reservoir maturity [67,68]. The presence of diamondoids is considered to be closely related to the geological maturity of an oil field [69-72]. Some investigators have used the relative abundance of diamondoids to fingerprint, identify and estimate the degree of oil cracking in an underground oil reservoir [73].

The relatively high melting points of diamondoids have caused their precipitation in oil wells, transport pipelines and processing equipment during production, transportation and refining of diamondoids-containing petroleum crude oil and natural gas [74-76]. This may cause fouling of pipelines and other oil processing facilities. Diamondoids deposition and possible fouling problems are usually associated with deep natural underground petroleum reservoirs that are rather hot and at high pressure. Other hydrocarbons with molecular weights in the same range as diamondoids are generally less stable and they crack at high temperatures [74].

The practice of petroleum production may lead to an environment that favors the reduction of diamondoids solubility in petroleum, their separation from petroleum fluid phase and their precipitation. For instance, it has been observed that phase segregation of diamondoids from the so-called "dry petroleum" (meaning petroleum fluids low in light hydrocarbons) streams takes place upon reduction of pressure and/or temperature of the stream. It has also been observed that in "wet petroleum" streams diamondoids partition themselves among the existing phases (vapor, liquid, solid) [74-76]. As a result, diamondoids may nucleate out of solution upon drastic changes of pressure and / temperature. Instabilities of this sort in crude oils may potentially initiate a sudden precipitation of other heavy organic compounds (including asphaltenes, paraffins and resins) on such nuclei [74-76]. Therefore, knowledge about solubility behavior of diamondoids in organic solvents and dense gases, as reported above, becomes important. It should be pointed out that among heavy organics in crude oil, asphaltenes, paraffins and diamondoids are propounded in nanotechnology subjects. By using the proper solvents, asphaltene molecules can form nanometer-sized micelles and micelle self-assembly (coacervate) particles [77].

Adamantane and diamantane are usually the dominant diamondoids found in petroleum and natural gas pipeline deposits [74,75]. This is because diamondoids are soluble in light hydrocarbons at high pressures and temperatures. Upon expansion of the petroleum fluid coming out of the underground reservoir and a drop of its temperature and pressure, diamondoids could deposit.

Deposition of adamantane from petroleum streams is associated with phase transitions resulting from changes in temperature, pressure, and/or composition of reservoir fluid. Generally, these phase transitions result in a solid phase from a gas or a liquid petroleum fluid. Deposition problems are particularly cumbersome when the fluid stream is dry (i.e. low LPG content in the stream). Phase segregation of solids takes place when the fluid is cooled and/or depressurized. In a wet reservoir fluid (i.e. high LPG content in the stream) the diamondoids partition into the LPG-rich phase and the gas phase. Deposition of diamondoids from a wet reservoir fluid is not as problematic as in the case of dry streams [74,75].

**Separation, Detection and Measurement:** Petroleum crude oil, gas condensate and natural gas are generally complex mixtures of various hydrocarbons and non-hydrocarbons with diverse molecular weights. In order to analyze the contents of a petroleum fluid it is a general practice to separate it first to five basic fractions namely, VOLATILES, SATURATES, AROMATICS, RESINS and ASPHALTENES [74,77]. Volatiles consist of the low-boiling fraction of crude separated in room temperature and under vacuum ($\sim$ 27°C and 10 mm-Hg) from crude oil. The contents of the volatiles fraction can be further analyzed using gas chromatography. The remaining four fractions are separated from the vacuum residue with the use of column liquid chromatography (i.e. SARA separation [78]). Initially the asphaltenes-fraction of the sample is removed by the ASTM D3279-90 separation method [79]. Then the saturates-fraction is extracted with n-hexane solvent by passing the sample through a liquid chromatography column that is packed with silica gel and alumina powder. Diamondoids are saturated hydrocarbons. Therefore analysis to determine their presence in a crude oil must be performed on the saturates-fraction. Mass spectrometry analyses must then be performed on the low-boiling part of the saturates-fraction to determine whether diamondoids are present in this petroleum fluid. To achieve this, the saturates-fraction is analyzed through gas chromatography / mass spectrometry (GC/MS) [11,74,75,80].

After GC/MS analyzer is informed of the standard molecular fragmentation spectrums of diamondoids (like Figure 14), diamondoids molecular weights are given to the GS/MS. Some representative chromatograms of petroleum fluids containing diamondoids obtained by various investigators [11,74,75,81] from the GC/MS are shown in Figures 15-18. Adamantane, if present in the sample, will elute between $nC_{10}$ and $nC_{11}$; diamantane will elute between $nC_{15}$ and $nC_{16}$; triamantane will elute between $nC_{19}$ and $nC_{20}$, etc. [75]. For example, the unknown peak with retention time of 5.70 eluted between $nC_{15}$ and $nC_{16}$ in Figure 15 is indicative of the probable existence of diamantane in the sample. A full scan total ion chromatogram (TIC) of the sample is turned around from 100 m/z to 1000 m/z using molecular fragmentation spectrums (like Figure 14 for adamantane). With the availability of standard molecular fragmentation spectrums of diamondoids (like Figure 14 for adamantane) [81,82] diamondoids are identified in the sample.

There exist a number of other methods for the separation of diamondoids from petroleum fluids or natural gas streams: (i) A gradient thermal diffusion process [54] is proposed for separation of diamondoids. (ii). A number of extraction and absorption methods [53,83] have been recommended for removing diamondoid compounds from natural gas streams. (iii). Separation of certain diamondoids from petroleum fluids has been achieved using zeolites [56,84] and a number of other solid adsorbents.

## Diamondoids as Templates and Molecular Building Block

Each successively higher diamondoid family shows increasing structural complexity and varieties of molecular geometries. Sui generis properties of diamondoids have provoked an extensive range of inquiries in different fields of science and technology. They have been used as templates for crystallization of zeolite catalysts [85], in the synthesis of high-temperature polymers [86], in drug delivery and drug targeting, in nanotechnology, in DNA directed assembly, in DNA-protein nanostructures and in host-guest chemistry.

**In Polymers:** Diamondoids, especially adamantane, its derivatives and diamantane, can be used for improvement of thermal stability and other physicochemical properties of polymers and preparation of thermosetting resins, which are stable at high temperatures. For example, it is demonstrated that acetylene groups on highly hindered diamondoid cage compounds (like 1,3-Diethynyladamantane) can be polymerized thermally to give thermoset polymers. The resulting thermoset polymer was shown to be stable to 475 $^{o}$C in air and exhibited less than 5% weight loss in air after 100 h at 301 $^{o}$C. This unusually high thermal stability for an aliphatic hydrocarbon polymer obviously results from the presence of adamantane units in the polymer backbone which due to their "diamond-like" structure, retards degradation reactions resulting from either nucleophilic or electrophilic attack, or from elimination reactions. It is also demonstrated that synthesis of acetylenic derivatives of diamantane and their thermal polymerization give materials with thermooxidative stabilities significantly greater than those of the corresponding adamantane polymers. The presence of diamantyl groups enhances the thermal stability of the resulting thermoset resins by nearly 50 $^{o}$C over the corresponding adamantane-based polymers [87]. Another example of high temperature polymers of this category is polymerization of diethynyl diamantane [88]. Adamantyl-substituted poly(m-phenylene) is synthesized starting with 1,3-dichloro-5- (1- adamantyl) benzene monomers (Figure 19) and it is also shown to have a high degree of polymerization and stability, decomposing at high temperatures of around 350°C [89].

Diamantane-based polymers are synthesized to take advantage of their stiffness, chemical and thermal stability, high glass transition temperature, improved solubility in organic solvents and retention of their physical properties at high temperatures. All these special properties result from their diamantane-based molecular structure [90]. Polyamides are high-temperature polymers with a broad range of applications in different scientific and industrial fields. However, their process is very difficult because of poor solubility and lack of adequate thermal stability retention [90]. Incorporation of 1,6- or 4,9-diamantylene groups into the polyamides results in improvement in their thermal stability and satisfactory retention of their storage modulus (~ stress/strain) above 350˚C [90]. These characteristics as well as improved viscosity and solubility in organic solvents, are specific to polyamides which are derived from 4,9-bis(4-

aminophenyl)diamantane and 4,9-bis[4-(4-aminophenoxy)phenyl] diamantane (Figure 20).

Star polymers are a class of polymers with interesting rheological and physical properties. The tetra-functionalized adamantane cores (adamantyls) have been employed as initiators in the atom transfer radical polymerization (ATRP) method applied to the styrene and various acrylate monomers (see Figure 21). As a result of this process star-like polymers have been produced with a wide range of molecular weights [91].

In another study, the introduction of an adamantyl group to the poly (etherimide) structure caused polymer glass transition temperature, Tg , and solubility enhancements in some solvents like chloroform and other aprotic solvents [92].

The introduction of bulky side-chains which contain adamantyl group to the poly(p-phenylenevinylene) (PPV), a semiconducting conjugated polymer, decreseases the number of interchain interactions. This action will reduce the  , aggregation quenching and polymer photoluminescence properties would be improved [93].

Substitution of the bulky adamantyl group on the C(10) position of  the biliverdin pigments structure leads to the distortion of helical conformation and hence the pigment color would shift from blue to red [94].

A 3-D four-fold interpenetrating coordination diamondoid polymer framework with adjustable porous structure was synthesized recently.  It was demonstrated that this polymer framework is capable of trapping gaseous molecules and may be useful for gas storage [95].

**In Nanotechnology:** Nanotechnology is the branch of engineering that deals with the manipulation of individual atoms, molecules and systems smaller than 100 nanometers. Two different methods are envisioned for nanotechnology to build nanostructured systems, components and materials: One method is named the "top-down" approach and the other method is called the "bottom-up" approach.  In the top-down approach the idea is to miniaturize the macroscopic structures, components and systems towards a nanoscale of the same.  In the bottom-up approach the atoms and molecules constituting the building blocks are the starting point to build the desired nanostructure [96-98].

Various illustrations are available in the literature depicting the comparison of top-down and bottom-up approaches [96,97].   In the top-down method a macro-sized material is reduced in size to reach the nanoscale dimensions. The photolithography used in semiconductor industry is an example of the top-down approach. The bottom-up nanotechnology is the engineered manipulation of atoms and molecules in a user-defined and repeatable manner to build objects with certain desired properties.  To achieve this goal a number of molecules are identified as the molecular building blocks (MBBs) of nanotechnology among which diamondoids are the most important ones owing to their unique properties [6, 99-104]. Diamondoids can be divided into two major clusters based

_______________________________

upon their size: lower diamondoids (1-2 nm in diameter) and higher diamondoids (>2 nm in diameter).

The building blocks of all materials in any phase are atoms and molecules. Their arrangements and how they interact with one another define many properties of the material. The nanotechnology MBBs, because of their sizes of a few nm's, impart to the nanostructures created from them new and possibly preferred properties and characteristics heretofore unavailable in conventional materials and devices. These nanosize building blocks are intermediate in size lying between atoms and microscopic and macroscopic systems. These building blocks contain a limited and countable number of atoms. They constitute the basis of our entry into new realms of bottom-up nanotechnology [97,98].

The controlled and directed organization of molecular building blocks (MBBs) and their subsequent assembly into nanostructures is one fundamental theme of bottom-up nanotechnology. Such an organization can be in the form of association, aggregation, arrangement, or synthesis of MBBs through non-covalent van der Waals forces, hydrogen bonding, attractive intermolecular polar interactions, electrostatic interactions, hydrophobic effects, etc [97].

The ultimate goal of assemblies of nanoscale molecular building blocks is to create nanostructures with improved properties and functionality heretofore unavailable to conventional materials and devices. As a result one should be able to alter and engineer materials with desired properties. For example, ceramics and metals produced through controlled consolidation of their MBBs are shown to possess properties substantially improved and different from materials with coarse microstructures. Such different and improved properties include greater hardness and higher yield strength in the case of metals and better ductility in the case of ceramic materials [102].

Considering that nanoparticles have much higher specific surface areas, in their assembled forms there are large areas of interfaces. One needs to know in detail not only the structures of these interfaces, but also their local chemistries and the effects of segregation and interaction between MBBs and their surroundings. The knowledge of ways to control nanostructure sizes, size distributions, compositions, and assemblies are important aspects of the bottom-up nanotechnology [97].

In general, nanotechnology molecular building blocks (MBBs) are distinguished for their unique properties. They include, for example, graphite, fullerene molecules made of various numbers of carbon atoms ($C60$, $C70$, $C76$, $C240$, etc.), carbon nanotubes, nanowires, nanocrystals, amino acids and diamondoids [97]. All these molecular building blocks are candidates for various applications in nanotechnology.

One of the properties used to distinguish MBBs from one another, is the number of their available linking groups. MBBs with three linking groups, like graphite, could only produce planar or tubular structures. MBBs with four linking groups may form three dimensional diamond lattices. MBBs with five linking groups may create 3-dimensional solids and hexagonal planes. The ultimate possibility is presently MBBs with six or more linking groups. Adamantane with six linking groups (see Figure 22) and higher diamondoids are of the latter category, which can construct many complex three-dimensional structures [102]. Such MBBs can have numerous applications in nanotechnology and they are of major interest in designing shape-targeted nanostructures including synthesis of supramolecules with manipulated architectures [105-109].

In addition to possessing six or more linking groups diamondoids have high strength, toughness, and stiffness when compared to other known MBBs. They are tetrahedrally symmetric stiff hydrocarbons. Strain-free structures of diamondoids give them high molecular rigidity, which is quite important for a MBB. High density, low surface energy, and oxidation stability are some other preferred properties of diamondoids as MBBs. Diamondoids have noticeable electronic properties [110]. In fact, they are H-terminated diamond and the only semiconductors, which show a negative electron affinity [6].

Since diamondoids possess the capability for derivatization they can be utilized for reaching to suitable molecular geometries needed for molecular building blocks of nanotechnology. Functionalization by different groups can produce appropriate reactants for desired reactions, microelectronics, optics, employing polymers, films, and crystal engineering.

Over 20,000 variants of diamondoids have been identified and synthesized and even more is possible [102], providing a rich and well-studied set of MBBs.

Adamantane can be used in molecular studies and preparation of fluorescent molecular probes [111]. Because of its incomparable geometric structure, the adamantane core (adamantyl) can impede interactions of fluorophore groups and self-quenching would diminish due to steric hindrance. Hence, mutual quenching would be diminished, allowing the introduction of several fluorescent groups to the same molecular probe in order to amplify their signals. Figure 23 shows the general scheme of an adamantane molecule with three fluorophore groups (F1) and a targeting group for attachment of biomolecules. Such a molecular probe can be very useful in DNA probing and especially in fluorescent-in-situ hybridization (FISH) diagnostics [112].

Due to their demanding synthesis, diamondoids are helpful models to study structure-activity relationships in carbocations and radicals, to develop empirical computational methods for hydrocarbons and to investigate orientational disorders in molecular crystals as well [5,32].

Atomic force microscope (AFM) is a powerful nanotechnology tool for molecular imaging and manipulations. One major factor limiting resolution in AFM to observe individual biomolecules such as DNA is the low sharpness of the AFM tip that scans the sample. Nanoscale 1,3,5,7-Tetrasubstituted adamantane is found to serve as molecular tip for AFM and may also find application as chemically well-defined objects for calibration of commercial AFM tips [113].

One of the branches of nanotechnology is called "crystal engineering". Crystal engineering is a new concept through, which the power of non-covalent intermolecular forces is used in the solid-state to design new nanomaterials with desired functions.

The approach in crystal engineering is to learn from known crystalline structures of, for example, minerals in order to design compounds with desired properties. Crystal engineering is considered to be a key new technology with applications in pharmaceuticals, catalysis, and materials science. The structures of adamantane and other diamondoids have received considerable attention in crystal engineering due to their molecular stiffness, derivatization capabilities and their six or more linking groups [114-117].

A concept named "molecular manufacturing" which was originally proposed by K. Eric Drexler [99] in 1992 has attracted the attention of some investigators [100,118-121]. Molecular manufacturing is defined as "the production of complex structures via non-biological mechanosynthesis (and subsequent assembly operations)" [99]. A chemical synthesis controlled by mechanical systems operating on atomic-scale and performing direct positional selection of reaction sites by atomic-precision manipulation systems is known as mechanosynthesis.

Due to their strong stiff structures containing dense, three-dimensional networks of covalent bonds, are one of the favorite sets of molecules considered for molecular manufacturing as originally proposed, described and analyzed in "Nanosystems". Diamondoid materials, if they could be synthesized as proposed, could be quite strong but light.  For example, diamondoids are being considered to build stronger, but lighter, rockets and other space components and a variety of other earth-bound articles for which the combination of weight and strength is a consideration [99,100,121-123].

Some of the applications of molecular manufacturing based on diamondoids are: the design of an artificial red blood cell called respirocyte, nanomotors, nanogears, molecular machines and nanorobots [102,123-126]. The other potential application of molecular manufacturing of diamondoids is in the design of molecular capsules and cages for various applications including drug delivery.

For the concept of molecular manufacturing to become successful, a systematic study of the fundamental theory of the molecular processes involved and the possible technological and product capabilities are needed [127].

**In Drug Delivery and Drug Targeting:** The unique structure of adamantane is reflected in its highly unusual physical and chemical properties, which can have many applications including drug design and drug delivery. The carbon skeleton of adamantane comprises a cage structure, which may be used for the encapsulation of other compounds, like drugs.  Although adamantane has been the subject of many research projects in the field of pharmacophore-based drug design, its application to drug delivery and drug targeting systems is a new matter of considerable importance [128].

Diamondoids could also be used as encapsulated, cage-shaped molecules in designing drug delivery systems. If the drug doesn't accumulate in its exact sight of action, it would not be able to produce the intended therapeutic effects even by using it at a high concentration. The nanometer size of diamondoid molecules makes them possible to enter living cells while carrying the drug into the cells. Commonly particles with less than 100 nm size can enter the cells, whereas diamondoids are even smaller than 10 nm. Furthermore, due to their high stability and solubility in blood plasma they are expected to play an important part in the future drug delivery systems [128].

Furthermore, polymantanes have the potential to be utilized in the rational design of multifunctional drug systems and drug carriers. In host-guest chemistry there is plenty of room for working with diamondoids.

In pharmacology, two adamantane derivatives, Amantadine (1-adamantaneamine hydrochloride) and Rimantadine (α-methyl-1-adamantane methylamine hydrochloride) (see Figure 24) have been well known because of their antiviral activity [129]. The main application of these drugs is prophylaxis (treatment to prevent the onset of a particular disease) and treatment of influenza-A viral infections. They are also used in the treatment of Parkinsonism and inhibition of hepatitis-C virus. Memantine (1-amino-3,5-dimethyladamantane) (see Figure 24) has been reported effective in slowing the progression of Alzheimer's disease [130].

The site of action for Memantine is the central nervous system (CNS) and it has CNS affinity. Amantadine and Rimantadine can penetrate to the CNS and cause some adverse effects.
These observations lead one to come to this conclusion that the adamantane nucleus which is present in all of the aforesaid drugs might be responsible for their penetration to the BBB and their accumulation in the CNS. In other words, the adamantane nucleus is likely to possess a so-called intimate CNS tropism (because of its hydrophobicity) and cause these drugs to show such a CNS affinity [131].

Furthermore, because the half-life of Amantadine and Rimantadine in bloodstream is long (12–18 hours for Amantadine and 24-36 hours for Rimantadine in young adults), utilization of adamantane derivative carriers can probably prolong drug presence time in blood circulation.

Extensive investigations have been performed related to synthesis of new adamantane derivatives with better therapeutic actions and less adverse effects. For example, it has been proven that adamantylamino-pyrimidines & -pyridines are strong stimulants of tumor necrosis factor-α (TNF-α) [132]. TFN is referred to a substance that can improve the body's natural response to cancer by killing the cancer cells. Another example is 1,6-diaminodiamantane [87], which possesses an antitumor and antibacterial activity. Also, many derivatives of aminoadamantanes have antiviral activity like 3-(2-adamantyl) pyrolidines with two pharmacophoric amine groups which have antiviral activity against influenza-A virus [133].

Some derivatives of adamantane with antagonist or agonist effects have also been synthesized. For instance, monocationic and dicationic adamantane derivatives block the α-amino-3-hydroxy-5-methylisoxazole-4-propionic acid (AMPA) receptors, N-methyl-D-aspartate (NMDA) receptors [134-136] and 5-hydroxytryptamine (5-HT3) receptors [137].

The monocationic and dicationic adamantane derivatives have been employed to investigate the topography of the channel binding sites of AMPA and NMDA receptors [135].

A dicationic adamantane derivative has been exploited as a selective and specific marker of native AMPA receptor assembly to determine the distribution of AMPA receptor subtypes among populations of the rat brain cells [134,136]. Other examples include antagonism of 5- HT3 and agonism of 5-HT4 receptors by aza(nor)adamantanes [138], P2X7 receptors antagonism by adamantane amides [139], antagonism of voltage gated calcium channels and probably activation of γ-aminobutyric acid (GABA) receptors by an adamantane amine derivative that results in its anticonvulsive and antinociceptive actions [140], and inhibition of glucosylceramidase enzyme and glycolipid biosynthesis by a deoxynojirimycin-bearing adamantane derivative leading to the strong anti-inflammatory and immunosuppressive activities [141].

Attaching some short peptidic sequences to adamantane makes it possible to design novel antagonists. The bradykinin antagonist, which is used as an anti-cancer agents is an example. The adamantane-based peptidic bradykinin analog was utilized in structure-activity relationship (SAR) studies on the bradykinin receptors and showed a potent activity in inhibition of bradykinin-induced cytokines release and stimulation of histamine release [142].

In an attempt to design the β-turn-peptide-mimics, aspartic acid (an amino acid also known as aspartate) and lysine (an amino acid especially found in gelatin and casein) were attached to each amine group of 1,3-diaminoadamantane in the form of amide bonds. The term β–turn refers to a peptide chain which forms a tight loop such that the carbonyl oxygen of one residue is hydrogen bonded with the amide nitrogen of a residue located three positions down the chain. The β-turn-peptide-mimics display some degree of fibrinogen-GPIIB/IIIA antagonism [143].

Adamantane derivatives can be employed as carriers for drug delivery and targeting systems. Due to their high lipophilicity, attachment of such groups to drugs with low hydrophobicity could lead to a substantial increase of drug solubility in lipidic membranes and thus increases of its uptake.

Furthermore, the large number of linking group possibilities of adamantane and other diamondoids (six or more) makes it possible to introduce several functional groups consisting of drug, targeting part, linkers, etc to them without undesirable interactions. In fact, adamantane derivatives can act as a central core for such drug systems.

_______________________

As an example, short peptidic sequences can be bound to adamantane and provide a binding site for connection of macromolecular drugs (like proteins, nucleic acids, lipids, polysaccharides, etc.) as well as small molecules. Hence, short amino acid sequences can have linker roles, which are capable of drug release in the target site. There are some successful examples of adamantyl (adamantane core) application for delivery of drugs to the brain [131]. For this purpose, 1-adamantyl was attached to several AZT (Azidothymidine) drugs via an ester spacer and these prodrugs could pass the blood-brain barrier (BBB) easily to reach the brain. The drugs concentration after using such lipophilized prodrugs was measured in the brain tissue and showed an increase of 7-18 fold in comparison with AZT drugs without adamantane vector. The ester spacer link is resistant to the plasma esterases, but it is cleaved after passing through the BBB by brain tissue esterases. Overall, it is important to note that adamantane is now considered a successful brain-directing drug carrier.

Another example of adamantane utilization for poorly-absorbed-drug delivery to the brain is the conjugation of [D-Ala2]Leu-enkephalin derivatives with a 1-adamantyl [144]. The antinociceptive effect of Leu-enkephalin disappears when it is administered peripherally since proteolytic enzymes would decompose it.   As a result it cannot penetrate into the CNS (Central Nervous System). It is feasible to conjugate the [D-Ala2]Leu-enkephalin with a 1-adamantane vector via an ester, amide or a carbamate linkage in order to enhance the drug lipophilicity and thus facilitate its delivery across the blood-brain barrier (BBB) to the brain [144]. The adamantane-conjugated [D-Ala2]Leu-enkephalin prodrugs (Figure 25) are highly lipophilic and show a significant antinociceptive effect because of their ability to cross the BBB [132]. These results suggest that adamantyl is a promising brain-directing drug vector providing a high lipophilicity, low toxicity and high BBB permeability for sensitive and poorly absorbed drugs [144,145].

Adamantane has also been used for lipidic nucleic acid synthesis as a hydrophobic group. Two major problems in gene delivery are the low uptake of nucleic acids by cells and their instability in blood medium. An increase in lipophilicity using hydrophobic groups would probably lead to improvement of uptake and an increase in intracellular concentration of nucleic acids. In this case, an amide linker is used to attach the adamantane derivatives to a nucleic acid sequence. Such a nucleic acid derivatization has no significant effect on hybridization with the target RNA.  Lipidic nucleic acids possessing adamantane derivative groups can be also exploited for gene delivery [146].

Recently, synthesis of a polyamine adamantane derivative have been reported which has a special affinity for binding to the major grooves in double-stranded DNA [147]. It should be pointed out that most of the polyamines have affinity for binding to double-stranded RNA, thus making RNA stabilized. DNA selectivity is one of the outstanding features of the said ligand. This positive nitrogen-bearing ligand has a tendency to establish hydrophobic interactions with deeper DNA grooves due to its size and steric properties. Such an exclusive behavior occurs because the ligand fits better in the DNA major grooves. This bulky ligand size is the same as zinc-finger protein which also binds to DNA major grooves.

Higher affinity of adamantane-bearing ligand to DNA, instead of RNA, probably arises from the presence of adamantane and leads to DNA stabilization. This fact may be exploited for using such ligands as stabilizing carriers in gene delivery. Adamantane causes lipophilicity to increase as well as DNA stabilization. Furthermore, ligand/groove size-based targeting might also be possible with less specificity by changing the bulk and conformation of ligand.

Polymers conjugated with 1-adamantyl moieties as lipophilic pendent groups can be utilized to design nanoparticulate drug delivery systems. Polymer 1 in Figure 26, which is synthesized by homopolymerization of ethyladamantyl malolactonate, can be employed as highly hydrophobic blocks to construct polymeric drug carries. In contrast, polymer 2 (Figure 26), which is synthesized by copolymerization of polymer 1 with benzyl malolactonate, is water-soluble and its lateral carboxylic acid functions can be used to bind biologically active molecules in order to achieve targeting as well. These represent examples of the possibilities of producing adamantane-based pH-dependant hydrogels and intelligent polymeric systems [148].

**In DNA Directed Assembly and DNA-Protein Nanostructures:** Due to the ability of adamantane and its derivatives to attach to DNA, it is possible to construct well-defined nanostructures consisting of DNA fragments as linkers between adamantane cores.  This could be a powerful tool to design DNA-directed nanostructured self-assemblies [149].

A unique feature of such DNA-directed self-assemblies is its site-selective immobilization, which makes possible to construct well-defined nanostructures. On the other hand, the possibility of the introduction of a vast number of substitutes (like peptidic sequences, nucleoproteins, hydrophobic hydrocarbon chains, etc.) to an adamantane core (adamantyl) makes such a process capable of designing steric colloidal and supramolecular conformations via setting hydrophobic / hydrophilic and other interactions. In addition, the rigidity of adamantane structure can provide strength and rigidity to such self-assemblies [150].

Bifunctional adamantyl, as a hydrophobic central core, can be used to construct peptidic scaffolding [151] as shown in Figure 27. This is the reason why adamantane is considered as one of the best MBBs. This may be considered an effective and practical strategy to substitute different amino acids or DNA segments on the adamantane core (Figure 28). In other words, one may exploit nucleic acid (DNA or RNA) sequences as linkers and DNA hybridization (DNA probe) to attach to these modules with adamantane core.  Thus the DNA-Adamantane-Amino acid nanostructure may be produced.

The knowledge about protein folding and conformation in biological systems can be used to mimic the design of a desired nanostructure conformation from a particular MBB and to predict the ultimate conformation of the nanostructure [152].  Such biomimetic nano-assembly is generally performed step by step. This will allow observation of the effect of each new MBB on the nanostructure.  As a result it is possible to control accurate formation of the desired nanostructure. Biomimetic controlled and directed assembly can be utilized to investigate molecular interactions, molecular modeling and study of relationships between the composition of MBBs and the final conformation of the nanostructures. Immobilization of molecules on a surface could facilitate such studies [153].

Nucleic acid attachments to an adamantane core (adamantyl) can be achieved in several ways. At least two nucleic acid sequences, as linkage groups, are necessary for each adamantyl to form a nanostructured-self assembly. Various geometrical structures may be formed by changing the position of the two nucleic acid sequences with respect to each other on two of the adamantyl six adamantyl linking groups or by the addition of more nucleic acid sequences on the other linking groups. A number of alterations can be exerted on the nucleic acid sequences utilizing the new techniques developed in solid-phase genetic engineering for immobilized DNA alteration [154]. For instance, ligated DNA (enzymatically joined two linear DNA fragments through covalent bonds) may be employed to join the adamantyl nanomodules similar to what has been done on immobilized DNA in the case of gene assembly [154], provided some essential requirements could be met for the retention of enzyme activities. Instead of ligated DNA, one may use hybridized DNA as well. It may be possible to modify the amino acid parts, adamantane cores and DNA sequences of the resulting nanostructure. For example, using some unnatural (synthetic) amino acids [155,156] with appropriate folding characteristics, the ability of conformation fine-tuning could be improved. Hence, the assembling and composing of adamantyls as central cores, DNA sequences as linkers and amino acid substituents (on the adamantane) as conformation controllers may lead to the design of DNA-adamantane-Amino acid nanostructures with desired and predictable properties.

By all accounts, the hypothesis of formation of DNA+adamantane+amino acid nanoarchitectures is currently immature and amenable to many technical modifications. Advancement in this subject requires a challenging combination of state-of-the-art approaches of organic chemistry, biochemistry, proteomics and surface science.

A dendrimer-based approach for the design of globular protein mimics using glutamic (Glu) and aspartic (Asp) acids as building blocks has been developed [151]. The pre-assembled Glu/Asp dendrones were attached to a 1,3- bifunctional adamantyl based upon a convergent dendrimer synthesis strategy (see Figure 28). Three successive generations of dendrimers composed of an adamantane central core and two, six and fourteen chiral centers (all L-type amino acids) and thus four, eight and sixteen peripheral carbomethoxy groups, respectively, were synthesized. The adamantane core was selected to render the dendrimer structures spherical with the capability of different ligands incorporated into their peripheral reactive arms. The Glu dendritic scaffoldings 2b and 3b (Figure 28) showed a noticeable feature from the solubility standpoint. The resulting dendrimers dissolved slowly in warm water to form a clear solution in spite of the fact that Glu dendrones, on their own, are quite insoluble in water. This solubility enhancement probably results from the double layer structures of dendritic scaffoldings 2b and 3b in which a hydrophilic outer shell encircles the hydrophobic adamantane core. In other words, this property can be attributed to the role of the adamantane core due to its ability to bring about a wide range of changes in physicochemical properties and turn a completely hydrophobic molecule to a hydrophilic one.

**In Host-Guest Chemistry:** The main aim in host-guest chemistry is to construct molecular receptors by a self-assembly process so that such receptors could, to some extent, gain molecular recognition capability. The goal of such molecular recognition capability is to either mimick or blocks a biological effect caused by molecular interactions [157].

Calixarenes, which are macrocyclic compounds, are one of the best building blocks to design molecular hosts in supramolecular chemistry [158]. Synthesis of Calix[4]arenes which have been adamantylated has been reported [105,109]. In Calix[4]arenes, adamantane or its ester/carboxylic acid derivatives were introduced as substituents (Figure 29). The purpose of this synthesis was to learn how to employ the flexible chemistry of adamantane in order to construct different kinds of molecular hosts. The X-ray structure analysis of p-(1-adamantyl) thiacalix[4]arene [109] demonstrated that it contained four $CHCl_3$ molecules, one of which was located inside the host molecule cavity, and the host molecule assumed the cone-like conformational shape (Figure 30).

Some other types of macrocycle compounds have been synthesized using adamantane and its derivatives. Recently, a new class of cyclobisamides has been synthesized using adamantane derivatives, which shows the general profiles of amino acid (serine or cystine)-ether composites. They were shown to be efficient ion transporters (especially for Na+ ions) in the model membranes [159]. Another interesting family of compounds to which adamantane derivatives have been introduced in order to obtain cyclic frameworks is "crown ethers" [160]. The outstanding feature of these adamantane-bearing crown ethers (which are also called "Diamond Crowns") is that α-amino acids can be incorporated to the adamantano-crown backbone [160]. This family of compounds provides the valuable models for studying selective host-guest chemistry, ion transportations and ion-complexation [160].

Adamantane has been also used as a cage-like alicyclic (both aliphatic and cyclic) bridge to construct a new class of tyrosine-based cyclodepsipeptides (tyrosinophanes) [161]. Macrocyclic peptides composed of an even number of D and L amino acids can self-assemble to form a tube through which ions and molecules can be transported across the lipid bilayers. Although they rarely exist in nature, they are synthesized to be employed in the host-guest studies (Figure 31-a) and to act as ion transporters in the model membranes (Figure 31-b,c) [161]. The adamantane-bridged, leucine-containing macrocycle 31-b shows a modest ability to transport $Na^+/K^+$ ions across the model membranes [161]. The adamantane-constrained macrocycle 31-c is also suitable for attachment of different functional groups to design artificial proteins [161]. The adamantane-containing cyclic peptides are efficient metal ion transporters and utilization of adamantane in such compounds improves their lipophilicity and thus membrane permeability [162]. A new class of norbornene-constrained cyclic peptides has been synthesized using adamantane as a second bridging ligand (Figure 32). The macrocycle 32-a is a specific ion transporter for monovalent cations while the cyclic peptide 32-b is able to transport both mono- and divalent cations across the model membranes [162].

Peptidic macrocycles are especially useful models for discovering protein folding mechanisms and designing novel peptide-made nanotubes as well as other biologically important molecules. These large cyclic peptides tend to fold in such a way that they can adopt a secondary structure like β-turns, β-sheets and helical motifs. A new series of double-helical cyclic peptides have been synthesized among them are the adamantane-constrained cystine-based cyclic trimers {cyclo (Adm-Cyst)3}. They have attracted a great deal of attention due to their figure eight-like helical topologies and special way of hydrogen binding and symmetries [163,164] (Figure 33). The cyclo (Adm-Cyst)3 molecule was able to transport K+ ions through the model membranes and it was a valuable model to study the mechanism of secondary structure formation in proteins [165].

Cyclodextrins (CDs) are inclusion compounds formed by enzymatic decomposition of starch to the cyclic oligosaccharides containing six to eight glucose units. Depending on the number of glucose units, there are three types of natural CDs, namely α, β, and γ consisting of six, seven and eight glucoses, respectively. The interior lining of the parent CDs' cavities is somehow hydrophobic. Adamantane is one of the best guests entrapped within the CDs' cavities [166-168]. Its noticeable association constant with CDs ($\sim 10^4$ to $10^5$ M$^{-1}$) denotes a high affinity to interact with a hydrophobic pocket of CDs, which is a valuable linking system to join different molecules together (see Figure 34). Interestingly, this system adsorbs and immobilizes molecules on a solid support and has been exploited to immobilize an adamantane-bearing polymer onto the surface of a β-CD-incorporated silica support [169]. In this case adamantane acts as a linker to attach a dextran-adamantane-COOH polymer to a solid support through a physical entrapment mechanism and thus contributes to formation of a stationary phase for chromatogaraphic purposes [169]. The aforementioned stationary phase could be readily prepared under mild conditions and is stable in aqueous media. It revealed some cation-exchange properties suggesting its application to the chromatography of proteins [169].

Covalent attachment of adamantane molecules is a key strategy to string them together and construct molecular rods. The McMurry coupling reaction was employed to obtain poly-adamantane molecular rods (see Figure 35) [170]. As another example, synthesis of tetrameric 1,3-adamantane and it's butyl derivative has been reported [171] (see Figure 36).

## Discussion:

Diamondoids are organic compounds (hydrocarbons) with unique structures and properties. This family of compounds (with over 20,000 variants) is one of the best candidates for templates and molecular building blocks for synthesis of high temperature polymers, nanotechnology, drug delivery, drug targeting, DNA directed assembly, DNA-protein nanostructure formation and in host-guest chemistry.

Some of their derivatives have been used as antiviral drugs. Due to their flexible chemistry, they can be exploited to design drug delivery systems and in molecular nanotechnology. In such systems, they can act as a central lipophilic core and different parts like targeting segments, linkers, spacers, or therapeutic agents can be attached to the said central nucleus. Their central core can be functionalized by peptidic and nucleic acid sequences and also by numerous important biomolecules.

Furthermore, some adamantane derivatives possess special affinity to bind to DNA, thereby stabilizing it. This is an essential feature for a gene vector. Some polymers have been synthesized using adamantane derivatives, the application of which is under investigation for drug delivery.

Adamantane can be used to construct peptidic scaffolding and synthesis of artificial proteins. It has been introduced into different types of synthetic peptidic macrocycles which are useful tools in peptide chemistry and stereochemistry studies and have many other applications as well. Introduction of amino acid-functionalized adamantane to the DNA nanostructures might lead to construction of DNA-adamantane-amino acid nanostructures with desirable stiffness and integrity. Diamondoids can be employed to construct molecular rods, cages and containers and also for utilization in different methods of self-assembly. In fact, through the development of self-assembly approaches and utilization of diamondoids in these processes, it would be possible to design and construct novel nanostructures for effective and specific carriers for each drug.

The phase transition boundaries (phase envelop) of adamantane need to be investigated and constructed. Predictable and diverse geometries are important features for molecular self-assembly and pharmacophore-based drug design. Incorporation of higher diamondoids in solid-state systems and polymers should provide high-temperature stability, a property already found in polymers synthesized from lower diamondoids.

Diamondoids offer the possibility of producing a variety of nanostructural shapes including molecular-scale components of machinery such as rotors, propellers, ratches, gears, toothed cogs, etc. We expect them to have the potential for even more possibilities for applications in molding and cavity formation characteristics due to their organic nature and their sublimation properties. The diverse geometries and possibility of six or more attachment sites (linking groups) in diamondoids provide an extraordinary potential for the production of shape-derivatives.

## Glossary

**5-HT3 receptor** = 5-Hydroxytryptamine receptor**:** A receptor for serotonin (a neurotransmitter) which activates a variety of second messenger signaling systems and through them indirectly regulates the function of ion channels.

**Adamantologs**: Diamondoids

**Adamantyl**: The tetra-functionalized adamantane core.

**AMPAR** = (α-amino-3-hydroxy-5-methyl-4-isoxazole propionic acid) Receptors**:** A non-NMDA-type ionotropic transmembrane receptor for glutamate that mediates fast synaptic transmission in the central nervous system.

**Antagonist**: Antagonist is defined as a compound which is able to bind to a drug or endogenous chemical's receptors and block them. Hence, it competes with the drug/endogenous chemical to occupy the receptors and inhibits the pharmacological effects of a drug/endogenous chemical to appear.

**Antinociceptive**: Analgesic; Painlessness

**Aprotic solvent**: A solvent that has no OH groups and therefore cannot donate a hydrogen bond.

**Association constant**: In host-guest chemistry between host (H) and guest (G) molecules is defined as $K=[H\text{-}G].[H]^{-1}.[G]^{-1}$.

**BBB =** Blood-brain barrier

**Carbocation** = carbonium ion: An ion with a positively charged carbon atom.

**CNS** = Central Nervous System

**Crystal engineering**: Utilization of non-covalent intermolecular forces in the solid-state to design new nanomaterials with desired functions.

**Cyclodextrin (CD)**: An inclusion cyclic oligosaccharide compound containing six to eight glucose units.

**Cytokines**: Cytokines are some non-specific water-soluble glycoproteins with a short half-life produced and secreted abruptly by white blood cells in response to an external stimulus, which act as chemical messengers between cells.

**DFT** = Density-Functional Theory

**Enantiomeric**: Optical antipode, Enantiomers are a pair of optical isomers

**Esterases**: Enzymes that catalyze the hydrolysis of an ester bond.

**Facies**: A rock or stratified body distinguished from others by its appearance or composition.

**FCC = Face –Centered Cubic**: The **FCC structure** is a close-packed **structure**

**Fibrinogen-GPIIB/IIIA**: A complex responsible for aggregation of platelets in blood stream

**FISH =** Fluorescent-in-situ hybridization**:** The method for utilizing fluorescently labeled DNA probes to detect or confirm gene or chromosome abnormalities that are generally beyond the resolution of routine cytogenetics.

**GABA Receptors**: Receptors for Gamma-aminobutyric acid. GABA is an amino acid which acts as an inhibitory neurotransmitter in CNS.

**Histamine**: A chemical present in cells throughout the body and released during an allergic reaction.

**LPG** = Gas Condensate: Liquid petroleum gas

**MBB** = Molecular Building Blocks

**McMurry coupling reaction**: The mechanism of the McMurry coupling reaction consists of two defined steps: the reductive dimerization of the aldehyde or ketone, and then subsequent deoxygenation of the 1,2-diolate intermediate, yielding the alkene.

**Mechanosynthesis**: Chemical synthesis controlled by mechanical systems operating on atomic-scale and performing direct positional selection of reaction sites by atomic-precision manipulation systems.

**NMDAR**: An ionotropic receptor for glutamate. It plays a critical role in synaptic plasticity mechanisms and thus is necessary for several types of learning and memory.

**NMDAR** = N-Methyl-D-Aspartate Receptor**:** An ionotropic receptor for glutamate. It plays a critical role in synaptic plasticity mechanisms and thus is necessary for several types of learning and memory.

**NMR** = Nuclear Magnetic Resonance

**Norbornene**: A bicyclic olefin.

**Pharmacophore**: A part within a drug molecule which is believed to play a major role in interaction of that drug with its target. There may be more than one pharmacophoric site within the chemical structure of a drug molecule.

**Polymantanes**: Diamondoids

**Prodrug**: Inactive precursor of a drug which is converted into its active form in the body by normal metabolic processes).

**Proteolytic enzymes**: An enzyme that catalyzes the breakdown of proteins into their building blocks, the amino acids.

**RNA** = Ribonucleic Acid: It is a molecule present in the cell of all living beings and is essential for the synthesis of proteins.

**TNF** = Tumor necrosis factor: TNFs are among the important cytokines playing the key role in activation and induction of some immune system's cells and cellular immunity processes responsible for proinflammatory and inflammatory response reactions as well.

**Synaptic transmission**: Transmission through the junction across which a nerve impulse passes from an axon terminal to a neuron, muscle cell, or gland cell.

**Tropism**: Response/reaction of an organism to an external stimulus shown as movement.

**Zinc-finger protein**: A DNA-binding protein, which contains a zinc atom.

# References

[1]  R. C. Fort, *Adamantane, The chemistry of diamond molecules* (Studies in organic Chem.; v. 5), Marcel Dekker, New York, **1976**.

[2]  G. A. Olah (Editor), *Cage Hydrocarbons*, J. Wiley & Sons, New York, NY, **1990**.

[3]  A. P. Marchand, *Science*, 3 January **2003**, 299, 52-53.

[4]  W. H. Bragg, W. L. Bragg, *Nature*, **1913**, 91, 554-556.

[5]  G. J. Kabo, A. V. Blokhin, M. B. Charapennikau, A. G. Kabo, V. M. Sevruk, *Thermochimica Acta,* **2000**, 345, 125-133.

[6]  J. E. Dahl, S. G. Liu SG, R. M. Carlson, *Science* **2003**, 299, 96-99.

[7]  G. A. Mansoori, L. Assoufid, T. F. George, G. Zhang, Measurement, Simulation and Prediction of Intermolecular Interactions and Structural Characterization of Organic Nanostructures. In *Proceed. of Conf. on Nanodevices & Systems, Nanotech 2003*; *February 23-2,* **2003,** *San Francisco, CA*.

[8]  G. R. Desiraju, *J. Molecular Structure* **1996**, 374,191-198.

[9]  R.C. Fort, P. v. R. Schleyer, *Chemical Rev.s,* **1964**, 64, 277-300.

[10]  T. Clark, T. Mc O. Knox, M.A. McKervey, H. Mackle, J.J. Rooney, *J. American Chem. Soc.,* **1979**, 101(9), 2404-2420.

[11]  Wingert, W S, *Fuel.* Jan. **1992,** 71, 37-43.

[12]  *Aldrich - Catalog Handbook of Fine Chemicals*, Milwaukee, Aldrich Chemical Co., Inc., **1992**, 30.

[13]  J. Reiser, E. McGregor, J. Jones, R. Enick, G. Holder, *Fluid Phase Equil.* **1996**, 117,160-167.

[14]  K. W. Hedberg, *Part II. Determination of some molecular structures by the method of electron diffraction. A. Adamantane,* PhD Dissertation, Chemistry, **1948**, Cal Tech.

[15]  R. C. Fort, *Adamantane, The Chemistry of Diamond Molecules* Marcell Dekker, New York, NY, **1976**.

[16]  R. H. Boyd, S.N. Sanwal, S. Shary-Tehrany, D. McNally, *J. Phys. Chem.*, **1971,** 75, 1264-1271.

[17]  A. S. Cullick, J. L. Magouirk, H. J. Ng, Paper # P1994.01,  Proceed. of the 73[rd] Annual GPA Convention, **1994,** Gas Processors Association, Tulsa, OK.

[18]  A. van Miltenburg, W. Poot, T. W. de Loos, *J. Chem. & Eng. Data,* Sep-Oct **2000,** 45 (5), 977-979; W. Poot, K. M. Kruger, T. W. de Loos, *J. Chem. Thermod., Apr.* **2003,** 35 (4), 591-604; W. Poot, T. W. de Loos,  *Fluid Phase Equil.,* July 30, **2004,** 221 (1-2), 165-174.

[19]  R. H. Boyd, S. N. Sanwal, S. Shary-Tehrany, D. McNally,  *J. Phys. Chem.,* **1971**, 75, 1264-1271.

[20]  R. S. Buttler, A. S. Carson, P. G. Laye, W. V. Steele, *J. Phys. Chem.*, **1971**, 3, 277-280.

[21]  M. Manson, N. Rapport, E. F. Westrum, Jr., *J. Am. Chem. Soc.,* **1970**, 92, 7296-7299.

[22]  E. E. Baroody, G. A. Carpenter, *Rpt. Naval Ordnance Systems Command Task No. 331-003/067-1/UR2402-001* for Naval Ordnance Station, Indian Head, MD, **1972**, 1-9.

[23]  E. F. Westrum, Jr.,  *J. Phys. Chem. Solids*, **1961**, 18, 83-85.

[24]  S.-S. Chang, E. F. Westrum, Jr.,  *J. Phys. Chem.*, **1960**, 64, 1547-1551.

[25]  E. F. Westrum, Jr., M. A. McKervey, J. T. S. Andrews, R. C. Fort, Jr., T. Clark, *J. Chem. Thermod.*, **1978**, 10(10),  959-965.

[26]  G. M. Spinella, J. T. S. Andrews, W. E. Bacon,  *J. Chem. Thermodyn.*, **1978**, 10, 1023-1032.

[27]  R. Jochems, H. Dekker, C. Mosselman, G. Somesen, *J. Chem. Thermodyn.*, **1982**, 14, 395-398.

[28]  W. K. Bratton, I. Szilard, C. A. Cupas, *J. Org. Chem*, **1967**, 32, 2019-2021.

[29]  T. Clark, T. M. O. Knox, M. A. McKervey, H. Mackle, J. J. Rooney,  *J. Am. Chem. Soc.*, **1975**, 97, 3835-3836.

[30]  W. V. Steele, I. Watt, *J. Chem. Thermodyn.*, **1977**, 9, 843.

[31]  P. O'Brien, T. E. Jenkins,  *Phys. Stat. Solida* (a), **1981**, 67, K161-162.

[32]  M. A. McKervey, *Tetrahedron* **1980**, 36, 971-992.

[33]  P. v. R. Schleyer, *Angewandte Chemie* Int'l Ed'n in English, 17 Dec **2003,** 8(7), 529–529 (Published Online)**.**

[34]   S-G Liu, J.E. Dahl, R.M.K. Carlson, *Nanotech 2004 Conf. Technical Program Abstract,* NSTI, Cambridge, MA.; See also: P. A. Cahill, *Tetrahedron Lett's*, **1990**, 31, 5417-5420.

[35]  V. S. Smith, A. S. Teja, *J. Chem. Eng. Data,* **1996**, *41,* 923-925.

[36]   T. Kraska, K. O. Leonhard, D. Tuma, G. M. Schneider, *J. Supercritical Fluids,* **2002**, 23, 209–224.

[37]  I. Swaid, D. Nickel, G. M. Schneider, *Fluid Phase Equilib*., **1985**, *21*, 95-112.

[38]  S. J. Park, T. Y. Kwak, G. A. Mansoori, *Int'l J. Thermophysics*, **1987**, 8, 449-471.

[39]  R. Hartono, G. A. Mansoori, A. Suwono, *Chem. Eng. Commum.*, **1999,** 173, 23-42**.**

[40]  S. Landa, V. Machacek,  M. Mzourek, M. Landa, *Chim. Ind.* (Publ. No. 506), **1933**.

[41]  V. Prelog, R. Seiwerth *Ber. Dtsch. Chem. Ges.,* **1941**, 74, 1644-1648.

[42]    M. Navratilova, K. Sporka, *Appl. Catalysis A, General,* **2000**, 203,127-132.

[43] P. v. R. Schleyer, *J. Am. Chem. Soc*., **1957,** 79, 3292.

[44]  P. v. R. Schleyer, M.M. Donaldson, *J. Am. Chem. Soc*., **1960,** 82,4645-4651

[45]   W. B. Jensen, *The Lewis acid-base concepts, an overview*, J. Wiley & Sons, New York, NY, **1980**.

[46]  C. Cupas, P. v. R. Schleyer, D. J. Trecker, *J. Am. Chem. Soc.,* **1965,** 87, 917-918.

[47]  A. Schwartz, *Synthetic and mechanistic studies in adamantane chemistry*, Educational Research Inst. British Columbia, CA, **1973**.

[48]    R. B Morland, *Heteroadamantanes, The improved synthesis and some reactions of 2-heteroadamantanes*, PhD Dissertation, Chemistry, **1976,** Kent State Univ.

[49] Z. Kafka, L. Vodicka, *Collec. Czechoslovak Chem. Communic*., **1990**, 55(8): 2043-2045; See also: H. Hopf, *Classics in Hydrocarbon Chemistry, Syntheses, Concepts, Perspective,* Wiley-VCH Verlag GmbH & Co, **2000**.

[50]    M. Shibuya, T. Taniguchi, M. Takahashi, K. Ogasawara, *Tetrahedron Lett.s* **2002**, 43, 4145 - 4147.

[51]  J. E. Dahl, R. M. Carlson, L. Shenggao, **2003**, *Science J.,* 299, 23-25.

[52]  W.J. King, SPE Paper #17761, *SPE Gas Tech. Symp. Proceed.*, **1988**, 469-490, Soc. Petroleum Eng. Int'l, Richardson, TX.

[53]  R.A. Alexander, C.E. Knight, D.D. Whitehurst, *U.S. Patent No. 4982049*, **1991**.

[54]  R.A. Alexander, C.E. Knight, D.D. Whitehurst, *U.S. Patent No. 4952747*, **1990**.

[55]  R.A. Alexander, C.E. Knight, *U.S. Patent No. 4952748*, **1990**.

[56]  R.A. Alexander, C.E. Knight, D.D. Whitehurst, *U.S. Patent No. 4952749*, **1990**.

[57]  J. E. Dahl, J. M. Moldowan, K. E. Peters, G. E. Claypool, M. A. Rooney, G. E. Michael, M. R. Mello, M. L. Kohnen, *Nature*, 6 May **1999**, 399, 54-57.

[58]  Schoell M., Carlson R.M., *Nature*, 1999, 399, 15-16.

[59]  K. E. Peters; C.M. Walters, J. M. Moldowan, **2005**. *The Biomarker Guide*, 2nd Edition, parts 1 & 2, Cambridge Univ. press, Cambridge, UK; See also: L. K. Schulz, A. Wilhelms, E. Rein, A. S. Steen., *Organic Geochem.,* **2001,** 32 (3), 365-375.
[60]  Brocks J.J., Summons R.E., "Sedimentary Hydrocarbons, Biomarkers for Early Life", *Treatise on Geochemistry*, 2003, 8, 63-115.

[61]  William J.A., Bjoroy M., Dolcater D.J., Winters J.A., *Organic Geochemistry*, 1986, 10, 451-461.

[62]  Wignert W.S., *Fuel*, 1992, 71, 37-43.

[63]  Petrov A., Arefjev O.A., Yakubson Z.V., in *Advances in Organic Geochemistry*, Edition Techniq, Paris, 1974, 517-522.

[64]  R. Lin, Z. A. Wilk, *Fuel,* **1995,** 74(10), 1512-1521.

[65]  J. E. P. Dahl, J. M. Moldowan, T. M. Peakman, J. C. Cardy, E. Lobkovsky, M. M. Olmstead, P. W. May, T. J. Davis, J. W. Steeds, K. E. Peters, A. Pepper, A. Ekuan, R. M. K. Carlson, *Angew. Chem., Int. Ed.,* **2003**, 42, 2040-2044.

[66]  S. L. Richardson, T. Baruah, M. J. Mehl, M. R. Pederson, *Chem. Physics Lett.s,* **2005**, 403, 83–88; See also: Y. F. Chang, Y. L. Zhao, M. Zhao M, et al., *Acta Chimica Sinica,* Oct. 14, **2004,** 62 (19), 1867-1870.

[67]  K. Grice,  R. Alexander, R. I. Kagi, *Organic Geochem.*, **2000**, 31, 67-73.

[68]  L. K. Schulz, A Wilhelms, E Rein, AS Steen, *Org. Geochem*, **2001**, 32, 365-375.

[69]  J. H. Chen, J. M. Fu, G. Y. Sheng, D. H. Liu, J. J. Zheng, **1996**, *Org. Geochem.* 25, 170–190.

[70] Z.-N. Gao, Y.-Y. Chen and F. Niu, *Geochemical J.,* **2001,** 35, 155-168.

[71] A Shimoyama, H. Yabuta, *Geochemical J.,* **2002**, 36, 173-189.

[72] J. H. Chen, J. M. Fu, G. Y. Sheng, D. H. Liu, J. J. Zheng, *Org. Geochem.*, **1996**, 25, 170–190.

[73] S. A. Stout, G. A. Douglas, *Eniro.NVIRO. Forensics,* Dec. **2004,** 5(4), 225-235.

[74] D. Vazquez, J. Escobedo, G. A. Mansoori, *Characterization of Crude Oils from Southern Mexican Oilfields*. In *Proceed. of the EXITEP 98, Int'l Petroleum Tech.,* **1998,** *Exhibition, Placio de Los Deportes*; *15-18 November 1998; Mexico City, Mexico, D.F.*

[75] D. Vazquez, G. A. Mansoori, *J. Petroleum Sci. & Eng.* **2000**, 26,49-55.

[76] G. A. Mansoori, *J. Petroleum Sci. & Eng.* **1997**, 17,101-111.

[77] Priyanto S., Mansoori G.A., Aryadi S., *Chem. Eng. Sci.*, 2001, 56, 33–39.

[78] D. M. Jewell, E. W. Albaugh, B. E. Davis, R. G. Ruberto, **1974**, Ind. Eng. Chem. Fundam., 13 (3), 278-282.

[79] ASTM, *Book of ASTM Standards*, ASTM Int'l, West Conshohocken, PA, **2005**.

[80] R. Lin, Z. A. Wilk, *Fue*l, **1995,** 74(10), 1512-1521.

[81] APT, *GC and GC-MS analysis of NIGOGA reference samples (NSO-1 and JR-1),* 2 May, **2003**, Applied Petroleum Technology AS, Kjeller, Norway.

[82] A. Shimoyama, H. Yabuta, **2002**, *Geochemical J.*, 36, 173-189.

[83] J. K. Henderson and J. R. Sitzman, US Patent No. 000001185H, May 1993.

[84] L. D. Rollmann, L. A. Green, R. A. Bradway, H. K. C. Timken, *Catalysis Today,* **1996**, 31, 163-169.

[85] S. I. Zones, Y. Nakagawa, G. S. Lee, C. Y. Chen, L. T. Yuen, *Microporous & Mesoporous Materials,* **1998**, 21, 199-211.

[86] M. A. Meador, *Annu. Rev. Mater. Sci,* **1998**, 28, 599-630.

[87] A.A. Malik,' T. G. Archibald, and K. Baum¸ M. R. Unroe, *Macromolecules*, **1991**, 24, 5266-5268.

[88] C. Yaw-Terng, W. Jane-Jen, *Tetrahedron Lett.s,* **1995**, 36, 5805-5806.

[89]     L. J. Mathias, G. L. Tullos, *Polymer,* **1996**, 37, 3771-3774.

[90]     Chern Y-T, *Polymer* **1998**, 39,4123-4127; See also: Y.-T. Chern, W.-L. Wang, *Polymer,* **1998**, 39, 5501-5506

[91]     C.-F. Huang, H.-F. Lee, S.-W. Kuo, H. Xu, F.-C. Chang, *Polymer (Guilford),* **2004**, 45, 2261-2269.

[92]     G. C. Eastmond, M. Gibas, J. Paprotny, *European Polymer J.,* **1999**, 35, 2097-2106.

[93]     Y. K. Lee, H. Y. Jeong, K. M. Kim, J. C. Kim, H. Y. Choi, Y. D. Kwon, D. J. Choo, Y. R. Jang, K. H. Yoo, J. Jang, A. Talaie, *Current Appl. Physics,* **2002**, 2, 241-244; See also: H. Y. Jeong, Y. K. Lee, A. Talaie, K. M. Kim, Y. D. Kwon, Y. R. Jang, K. H. Yoo, D. J. Choo, J. Jang, *Thin Solid Films,* **2002**, 417, 171-174.

[94]     A. K. Kar, D. A. Lightner, *Tetrahedron, Asymmetry,* **1998**, 9, 3863-3880.

[95]     J. Zhang, W. B. Lin, Z. F. Chen, R. G. Xiong, B. F. Abrahams, H. K. Fun, *J. Chemical Soc.-Dalton Trans.,* **2001,** 12, 1806-1808.

[96]     R. W. Siegel, E. Hu, M. C. Roco, *Nanostructure Sci. & Tech. - A Worldwide Study. Prepared under the guidance of the IWGN, NSTC,* **1999,** WTEC, Loyola Collage in Maryland.

[97]     G. A. Mansoori, *Principles of Nanotechnology: Molecular Based Study of Condensed Matter in Small Systems*, World Scientific Pub. Co., New York, NY, **2005**.

[98]     G. A. Mansoori, T. F. George, G. Zhang, L. Assoufid (Ed's), *Molecular Building Blocks for Nanotechnology: From Diamondoids to Nanoscale Materials and Applications* (Topics in Applied Physics Series, Springer-Verlag), (to appear) **2006**.

[99]     K. E. Drexler, *Engines of Creation*. Anchor Press/Doubleday, Garden City, NY, **1986**.

[100]     K. E. Drexler, *Nanosystems, Molecular Machinery, Manufacturing, and Computation*, J. Wiley & Sons, New York, NY, **1992**.

[101]     G, C. McIntosh, M. Yoon, S. Berber, D. Tománek, *Physical Rev. B,* **2004**, 70, 045401.

[102]     R. C. Merkle, *Trends in Biotechnology* **1999**, 17, 271-274; See also: *ibid, Nanotechnology,* **2000**, 11, 89-99.

[103] Mansoori. G.A., "Advanced in Atomic and Molecular Nanotechnology", *United Nations Tech Monitor*, 2002, Special Issue, 53-59.

[104] Rawls R., *Material science*, 2002, 80, 13.

[105] V. Kovalev, E. Shokova, A. Khomich, Y. Luzikov, *New J. Chemistry,* April **1996,** 20(4), 483-492.

[106] D. N. CHIN, D. M. GORDON, G. M. WHITESIDES, Dec. 28 **1994,** *J. Amer. Chem. Soc.,*116(26), 12033-12044.

[107] O. Ermer, L. Lindenberg, *Helvetica Chimica Acta,* **1991,** 74(4), 825-877.

[108] O. Ermer, L. Lindenberg, *Helvetica Chimica Acta,* **1988,** 71(5), 1084-1093

[109] (a)-E. Shokova, V. Tafeenko, V. Kovalev, *Tetrahedron Lett.s,* **2002**, 43, 5153 – 5156; See also: (b)-K. A. Hirsch, S. R. Wilson, J. S. Moore, *Chemistry-A European J.,* **1997.** 3(5), 765-771.

[110] N. D. Drummond, A. J. Williamson, R. J. Needs, and G. Galli, *Physical Rev. Lett's,* **2005**, 95(9), 096801.

[111] P. R. Seidl, K. Z. Leal, *J. Molecular Structure: THEOCHEM,* **2001**, 539, 159-162.

[112] V. V. Martin, I. S. Alferiev, A. L. Weis, *Tetrahedron Lett.s,* **1999**, 40, 223-226.

[113] Q. Li, A. V. Rukavishnikov, P. A. Petukhov, T. O. Zaikova, and J. F. W. Keana, Org. Lett., Vol. 4, No. 21, 2002

[114] O. R. Evans, R.-G. Xiong, Z. Wang, G. K. Wong, W. Lin, *Angew. Chem. Int. Ed. Engl.,* **1999**, 38, 536-538; See also: O. R. Evans, W. B. Lin, *Chem. of Materials,* **2001,** 13(8), 2705-2712.

[115] R. Vaidhyanathan; S. Natarajan; C. N. R. Rao, *Encyclopedia of Supramolecular Chemistry,* **2004,** 1 – 12.

[116] M. J. Zaworotko, *Chem. Soc. Rev.s,* **1994**, 23(4), 283-288.

[117] D. S. Reddy, D. C. Craig, G. R. Desiraju, *J. Amer. Chem. Soc.,* **1996,** 118(17), 4090-4093.

[118] C. Phoenix, *Design of a Primitive Nanofactory, J. Evolution & Tech.,* **2003,** 13(2).

[119] R. A. Freitas Jr., R. C. Merkle, Kinematic *Self-Replicating Machines*, Landes Bioscience, Georgetown, TX, **2004.**

[120]  A. Herman, *Modelling Simul. Mater. Sci. Eng*., **1999,** 7 43-58.

[121] J. Peng, R. A. Freitas Jr., R. C. Merkle, J. Comp. Theor. Nanosci, 1 March **2004**, 62-70.

[122]  K. E. Drexler, C. Peterson, G. Pergamit, *Unbounding the Future, The Nanotechnology Revolution.* William Morrow & Co., New York, NY, **1991**.

[123]  K. E. Drexler, *Trends in Biotechnology,* **1999**, 17, 5-7.

[124]  K. Bogunia-Kubik, M. Sugisaka, *Biosystems,* **2002**, 65**,**123 - 138.

[125]  R. A. Freitas, *Nanotechnology,* **1996**, 2, 8-13.

[126]  R. A. Freitas, Jr., *Artifficial Cells, Blood Substitutes & Biotechnology,* **1998**, 26**,**411-430.

[127]  A. Herman, Modelling & Simulation in Materials Sci. & Eng., **1999,** 7(1), 43-58.

[128] Ramezani H., Mansoori G.A., "Diamondoids as Molecular Building Block for Nanotechnology, Drug Targeting and Gene delivery", to appear (2006).

[129] Anderzej O., *ActaBiochemical Polonica*, 2000, 47, 1-7.

[130]  J. G. Hardman, L. E. Limbird, *Goodman & Gilman's: The pharmacological basis of therapeutics.* 10th Edn. McGraw-Hill Book Co., New York, NY, **2001**.

[131]  N. Tsuzuki, T. Hama, M. Kawada, A. Hasui, R. Konishi, S. Shiwa, Y. Ochi, S. Futaki, K. Kitagawa, *J. Pharmaceutical Sci.,* **1994**, 83,481-484.

[132]  Z. Kazimierczuk, A. Gorska, T. Switaj, W. Lasek, *Bioorganic & Medicinal Chem. Lett.s,* **2001**, 11,1197-1200.

[133]  G. Stamatiou, A. Kolocouris, N. Kolocouris, G. Fytas, G. B. Foscolos, J. Neyts, E. De Clercq, *Bioorganic & Medicinal Chem. Lett.s,* **2001**, 11, 2137-2142.

[134]  M. V. Samoilova, S. L. Buldakova, V. S. Vorobjev, I. N. Sharonova, L. G. Magazanik, *Neuroscience,* **1999**, 94, 261-268.

[135]  K. V. Bolshakov, D. B. Tikhonov, V. E. Gmiro, L. G. Magazanik, *Neuroscience Lett.s,* **2000**, 291, 101-104.

[136]  S. L. Buldakova, V. S. Vorobjev, I. N. Sharonova, M. V. Samoilova, L. G. Magazanik, *Brain Research,* **1999**, 846, 52-58.

[137]  G. Rammes, R. Rupprecht, U. Ferrari, W. Zieglgansberger, C. G. Parsons, *Neuroscience Lett.s,* **2001**, 306, 81-84.

[138]  D. L. Flynn, D. P. Becker, D. P. Spangler, R. Nosal, G. W. Gullikson, C. Moummi, D.-C. Yang, *Bioorganic & Medicinal Chem. Lett.s,* **1992**, 2, 1613-1618.

[139]  A. Baxter, J. Bent, K. Bowers, M. Braddock, S. Brough, M. Fagura, M. Lawson, T. McInally, M. Mortimore, M. Robertson, *Chemistry Lett.s,* **2003**, 13, 4047-4050.

[140]  G. Zoidis, I. Papanastasiou, I. Dotsikas, A. Sandoval, R. G. Dos Santos, Z. Papadopoulou-Daifoti, A. Vamvakides, N. Kolocouris, R. Felix, *Bioorganic & Medicinal Chem.* **2005**, 13, 2791-2798.

[141]  C. Shen, D. Bullens, A. Kasran, P. Maerten, L. Boon, J. M. F. G. Aerts, G. van Assche, K. Geboes, P. Rutgeerts, J. L. Ceuppens, *Int'l Immunopharmacology,* **2004**, 4, 939-951.

[142]  S. Reissmann, F. Pineda, G. Vietinghoff, H. Werner, L. Gera, J. M. Stewart, I. Paegelow, *Peptides,* **2000**, 21, 527-533.

[143]  W. J. Hoekstra, J. B. Press, M. P. Bonner, P. Andrade-Gordon, P. M. Keane, K. A. Durkin, D. C. Liotta, K. H. Mayo, *Bioorganic & Medicinal Chem. Lett.s,* **1994**, 4, 1361-1364.

[144]  N. Tsuzuki, T. Hama, T. Hibi, R. Konishi, S. Futaki, K. Kitagawa, *Biochemical Pharmacology,* **1991**, 41, R5-R8.

[145]  K. Kitagawa, N. Mizobuchi, T. Hama, T. Hibi, R. Konishi, S. Futaki, *Chem. & Pharmaceutical Bulletin (Tokyo),* **1997**, 45, 1782-1787.

[146]  M. Manoharan, K. L. Tivel, P. D. Cook, *Tetrahedron Lett.s,* **1995**, 36, 3651-3654.

[147]  N. Lomadze, H. J. Schneider, *Tetrahedron Lett.s,* **2002**, 43, 4403 - 4405.

[148]  L. Moine, S. Cammas, C. Amiel, P. Guerin, B. Sebille, *Polymer,* **1997**, 38, 3121-3127.

[149]  I. Habus, Q. Zhao, S. Agrawal. *Bioconjugate Chem.,* **1995**, 6, 327-331; See also: M. Manoharan, K. L. Tivel, P. D. Cook, *Tetrahedron Lett.s,* **1995**, *36*, 3651-3654.

[150]  C. L. D. Gibb, X. Li, B. C. Gibb, **PNAS**, **2002**, 99(8), 4857-4862; See also: S. Aoki, M. Shiro, E. Kimura, *Chemistry - A European J.*, **2002,** 8(4), 929-939; See also: S. B. Copp, S. Subramanian, M. J. Zaworotko, *J. Amer. Chem. Soc.*, 1992, 114 (22), 8719-8720.

[151]  D. Ranganathan, S. Kurur, *Tetrahedron Lett.s,* **1997**, 38, 1265-1268.

[152]   A. K. Dillow, A. M. Lowman (Editors), *Biomimetic Materials & Design, Biointerfacial Strategies, Tissue Engineering, and Targeted Drug Delivery* Marcel Dekker, New York, NY, **2002**.

[153]   K. Busch, R. Tampé, *Rev.s in Molecular Biotechnology* **2001**, 82, 3-24.

[154]   J. H. Kim, J.-A. Hong, M. Yoon, M. Y. Yoon, H.-S. Jeong, H. J. Hwang, *J. Biotechnology,* **2002**, 96, 213 - 221.

[155]   C. J. Noren, S. J. Anthony-Cahil, M. C. Griffith, P. G. Schultz, *Science*, **1989,** 244, 182-188.

[156]   J. A. Piccirilli, T. Krauch, S. E. Moroney, S. A. Benner, *Nature*, **1990,** 343, 33-43.

[157]   Annon., *Host-Guest Molecular Interactions, from Chemistry to Biology*, CIBA Foundation Symposia Series, No. 158, J. Wiley & Sons, New York, NY, **1991**.

[158]   L. Mandolini, R. Ungaro (Editors), *Calixarenes in Action*, World Scientific Pub. Co., New York, NY, **2000**.

[159]   D. Ranganathan, M. P. Samant, R. Nagaraj, E. Bikshapathy, *Tetrahedron Lett.s,* **2002**, 43, 5145-5147.

[160]   D. Ranganathan, V. Haridas, I. L. Karle, *Tetrahedron,* **1999**, 55, 6643-6656.

[161]   D. Ranganathan, A. Thomas, V. Haridas, S. Kurur, K. P. Madhusudanan, R. Roy, A. C. Kunwar, A. V. Sarma, M. Vairamani, K. D. Sarma, *J. Organic Chem.,* **1999**, 64, 3620-3629.

[162]   D. Ranganathan, V. Haridas, S. Kurur, R. Nagaraj, E. Bikshapathy, A. C. Kunwar, A. V. Sarma, M. Vairamani, *J. Organic Chem.,* **2000**, 65, 365-374.

[163]   D. Ranganathan, V. Haridas, R. Nagaraj, I.L. Karle, L. Isabella, *J. Organic Chem.,* **2000**, 65, 4415-4422.

[164]   I. L. Karle, *J. Molecular Structure,* **1999**, 474, 103-112.

[165]   I. L. Karle, D. Ranganathan, *J. Molecular Structure,* **2003**, 647, 85-96.

[166]   C. Jaime, J. Redondo, F. Sanchez-Ferrando, A. Virgili, *J. Molecular Structure,* **1991**, 248, 317-329.

[167]   K. Fujita, W-H. Chen, D.-Q. Yuan, Y. Nogami, T. Koga, *Tetrahedron, Asymmetry,* **1999**, 10, 1689–1696.

[168]  D. Krois, U. H. Brinker, *J. Amer. Chem. Soc.*, **1998,** 120(45), 11627-11632.

[169]   C. Karakasyan, M.-C. Millot, C. Vidal-Madjar, *J. Chromatography B,* **2004**, 808, 63-67.

[170]   F. D. Ayres, S. I. Khan, O. L. Chapman, *Tetrahedron Lett.s,* **1994**, 35, 8561-8564.

[171]  T. Ishizone, H. Tajima, S. Matsuoka, S. Nakahama, *Tetrahedron Lett.s,* **2001**, 42, 8645-8647.

**Table I:** Some physical properties of diamondoids mostly compiled by ChevronTexaco

*www.moleculardiamond.com*

| Diamondoid Chemical Formula | Molecular Structure | Mw | MP [°C] | aBP [°C] | ρ [g/cc] | Crystal Structures |
|---|---|---|---|---|---|---|
| Adamantane $C_{10}H_{16}$ | 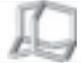 | 136.240 | 269. | 135.5 @ 10 *mm Hg* | 1.07 | Cubic, fcc |
| Diamantane $C_{14}H_{20}$ | 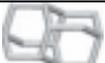 | 188.314 | 236.5 | 272 | 1.21 | Cubic, Pa₃ |
| Triamantane $C_{18}H_{24}$ | 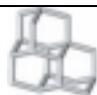 | 240.390 | 221.5 | 330 | 1.24 | Orthorhombic, Fddd |
| Tetramantanes $C_{22}H_{28}$ | 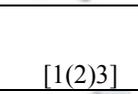 [1(2)3] | 292.466 | NA | NA | NA | NA |
| | 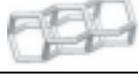 [121] | | 174 | NA | 1.27 | Monoclinic, P2₁/n |
| | 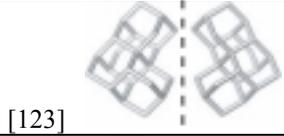 [123] | | NA. | NA | 1.32 | Triclinic, P1 |
| Pentamantanes $C_{26}H_{32}$ | 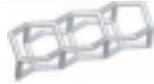 [1212] | 344.543 | NA | NA | 1.26 | Orthorhombic, P2₁2₁2₁ |
| | 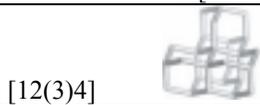 [12(3)4] | | NA | NA | NA | Monoclinic, P2₁/n |
| | 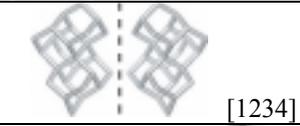 [1234] | | NA | NA | 1.30 | NA |
| | 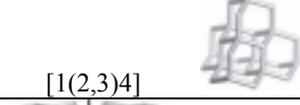 [1(2,3)4] | | NA | NA | 1.33 | Orthorhombic, Pnma |
| | 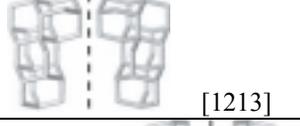 [1213] | | NA | NA | 1.36 | Triclinic, P-1 |
| | 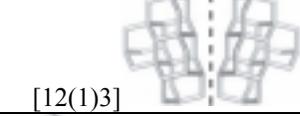 [12(1)3] | | NA | NA | NA | NA |
| Cyclohexamantane ( *peri-condensed* ) $C_{26}H_{30}$ | 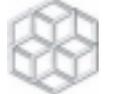 *Top* [12312] *Side* | 342.528 | >314 | NA | 1.38 | Orthorhombic, Pnma |
| Heptamantanes $C_{30}H_{34}$ | 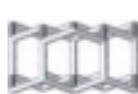 [121321] | 394.602 | NA | NA | 1.35 | Monoclinic, C2/m (#12) |
| | 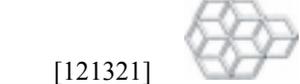 [123124] | 394.602 | NA | NA | NA | NA |

aBP=apparent boiling point, MP=melting point, Mw=molecular weight, $\rho$ =normal density

**Table II:** Vapor pressure equations of adamantane and diamantane for liquid-vapor and solid-vapor phase transitions

| Diamondoid | Phase Transition Kind | $ln$ P [kPa] = | Temp. Range [K] | Reference |
|---|---|---|---|---|
| Adamantane | liquid-vapor | -4670 / T + 14.75 | T > 543 K | [13] |
| | solid-vapor | - 6570/T + 18.18<br>- 6324.7/T + 17.827<br>- 9335.6/T + 65.206 - 15.349 $log$ T<br>- 7300/T + 31.583 - 4.376 $log$ T | 483 - 543<br>366 - 443<br>313 - 443<br>333 - 499 | [13]<br>[16]<br>[16]<br>[17] |
| Diamantane | liquid-vapor | -5680/T + 14.858 | 516 - 716 | [13] |
| | solid-vapor | -7330/T + 18.00<br>-7632.5/T + 18.333<br>-18981.3/T+190.735 -55.4418 $log$ T | 498 - 516<br>353 - 493<br>332 - 423 | [13]<br>[17]<br>[17] |

**Table III:** Thermodynamic properties of adamantane and diamantane. In this table the average of the values reported by various investigators are reported

| PROPERTY | Diamondoid | | Value | Units | T[K] | References |
|---|---|---|---|---|---|---|
| $\Delta H^{of}_{gas}$ | Adamantane | | -133.6 | kJ/gmol | | [10,19-21] |
| | Diamantane | | -145.9 | kJ/gmol | | [10] |
| $\Delta H^{of}_{solid}$ | Adamantane | | -191.1 | kJ/gmol | | [10,19-22] |
| | Diamantane. | | -241.9 | kJ/gmol | | [10] |
| $S^{o}_{solid}$ | Adamantane. (crystalline phase II) | | 195.83 | J/gmol•K | | [23,24] |
| | Diamantane. (crystalline phase III) | | 200.16 | J/gmol•K | | [25] |
| $C^{solid}_{P}$ (solid phase I) | Adamantane. | | 189.74 | J/gmol•K | 298.15 | [23] |
| | Diamantane. | | 220.2 | J/gmol•K | 295.56 | [26] |
| | | | 223.22 | | 298.15 | [25] |
| $\Delta H^{o}_{sublimation}$ | Adamantane. | | 59.9 | kJ/gmol | | [10,16,21,27,28] |
| | Diamantane. | | 96. | kJ/gmol | 305.-333 | [29] |
| $\Delta H_{phase.transition}$ | Adam. (Solid I-Solid II) | | 3.376 | kJ/gmol | 208.62 | [23,24] |
| | Diam. | Solid I-Solid III | 4.445 | kJ/gmol | 407.22 | [26] |
| | | Solid I-Solid II | 8.960 | kJ/gmol | 440.43 | |
| | | Solid I--Liquid | 8.646 | kJ/gmol | 517.92 | |
| $\Delta S_{phase.transition}$ | Adam. (Solid I-Solid II) | | 16.18 | J/mol•K | 208.62 | [26] |
| | Diam. | Solid I-Solid III | 10.92 | J/mol•K | 407.22 | |
| | | Solid I-Solid II | 20.34 | J/mol•K | 440.43 | |
| | | Solid I--Liquid | 16.69 | J/mol•K | 517.92 | |
| $\Delta H^{o}_{combustion}$ | Adam. (solid phase) | | -6030.04 | kJ/gmol | | [10,16,20,21] |
| | Diam. (solid phase) | | -8125.58 | kJ/gmol | | [29] |

**Note**: Solid adamantane possesses two crystalline phases and diamantane exists in three crystalline phases.

**Table IV:** Thermodynamic properties of methyl-derivatives of adamantane and diamantane. In this table the average of the values reported by various investigators are reported.

| DIAMONDOID | $\Delta H_{gas}^{of}$ [ kJ/gmol] | $\Delta H_{solid}^{of}$ [ kJ/gmol] | $\Delta H_{sublimation}^{o}$ [ kJ/gmol] | $\Delta H_{combustion}^{o}$ [ kJ/gmol] (solid phase) | References |
|---|---|---|---|---|---|
| ADAMMANTANE | - 133.6 | -191.1 | 59.9 | - 6030.04 | See Table II |
| 1-METHYL ADAMANTANE | - 171.6 | - 240.1 | 67.7 | - 6661.1 | [10,30] |
| 1,3-DIMETHYL ADAMANTANE | -219.0 | -287.3 | 67.8 | -7294.0 | [30] |
| 1,3,5-TRIMETHYL ADAMANTANE | -255.0 | -333.0 | 77.8 | -7927.4 | [30] |
| 1,3,5,7-TETRAMETHYL ADAMANTANE | -283.3 | - 370.7 | 82.4 | - 8568.7 | [10,30] |
| 2-METHYL ADAMANTANE | -151.7 | - 220.8 | 67.7 | - 6680.4 | [10,30] |
| DIAMANTANE | -145.9 | -241.9 | 96. | - 8125.6 | See Table II |
| 1-METHYL DIAMANTANE | -166.7 | -247.4 | 80.6 | - 8799.4 | [10] |
| 3-METHYL DIAMANTANE | -157.3 | -260.4 | 103.1 | -8786.36 | [10] |
| 4-METHYL DIAMANTANE | -182.1 | -261.5 | 79.4 | -8786.2 | [10] |

**Table V:** Solubilities of Diamondoids in Liquid Solvents at $25^O$ C. From [13].

| Solvent | Adamantane (wt%) | Diamantane (wt%) |
|---|---|---|
| Carbon tetrachloride | 7.0 | 5.0 |
| Pentane | 11.6 | 4.0 |
| Hexane | 10.8 | 3.9 |
| Heptane | 10.4 | 3.7 |
| Octane | 10.0 | 3.9 |
| Decane | 8.9 | 3.5 |
| Undecane | 7.9 | 3.2 |
| Tridecane | 7.3 | 2.7 |
| Tetradecane | 7.5 | 2.3 |
| Pentadecane | 7.1 | 2.2 |
| Cyclohexane | 11.1 | 6.3 |
| Benzene | 10.9 | 4.3 |
| Toluene | 9.9 | 4.5 |
| m-Xylene | 9.8 | 4.5 |
| p-Xylene | 9.6 | 4.5 |
| o-Xylene | 9.6 | 4.1 |
| THF | 12.0 | 4.0 |
| Diesel oil | 7.5 | 2.7 |
| 1,3, Dimethyl-adamantane | 6.0 | 2.0 |

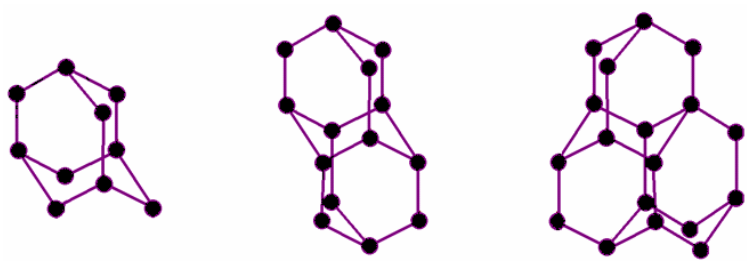

**Figure 1:** Molecular structures of adamantane, diamantane and trimantane, the smaller diamondoids, with chemical formulas $C_{10}H_{16}$, $C_{14}H_{20}$ and $C_{18}H_{24}$, respectively.

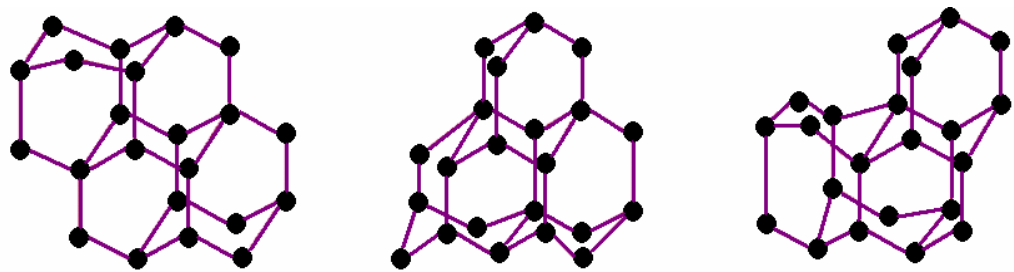

**Figure 2:** There are three possible tetramantanes all of which are isomeric, respectively from left to right as anti-, iso- and skew-tetramantane. Anti- and skew-tetramantanes, each, possess two quaternary carbon atoms, whereas iso-tetramantane has three quaternary carbon atoms. The number of diamondoid isomers increase appreciably after tetramantane.

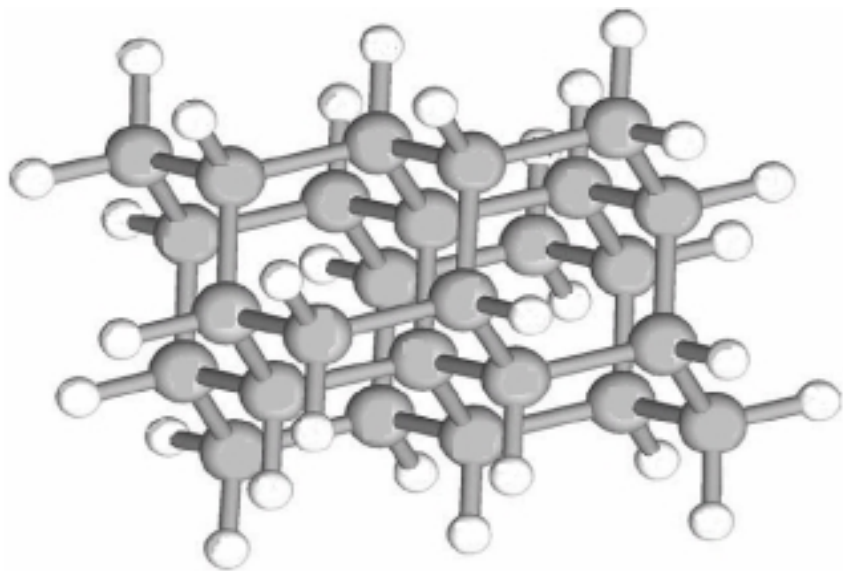

**Figure 3:**  Molecular structure of (*peri-condensed* ) cyclohexamantane (C$_{26}$H$_{30}$).  Darker spheres represent carbon atoms while lighter spheres are hydrogen atoms.

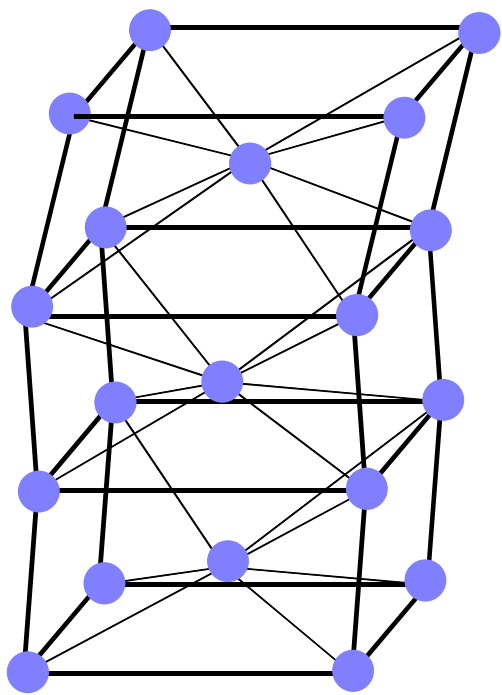

**Figure 4:** The quasi-cubic units of crystalline network for 1,3,5,7-tetrahydroxyadamantane. Molecules have been shown as blue spheres and hydrogen bonds as solid linking lines. This crystalline structure is similar to that of CsCl. From [8].

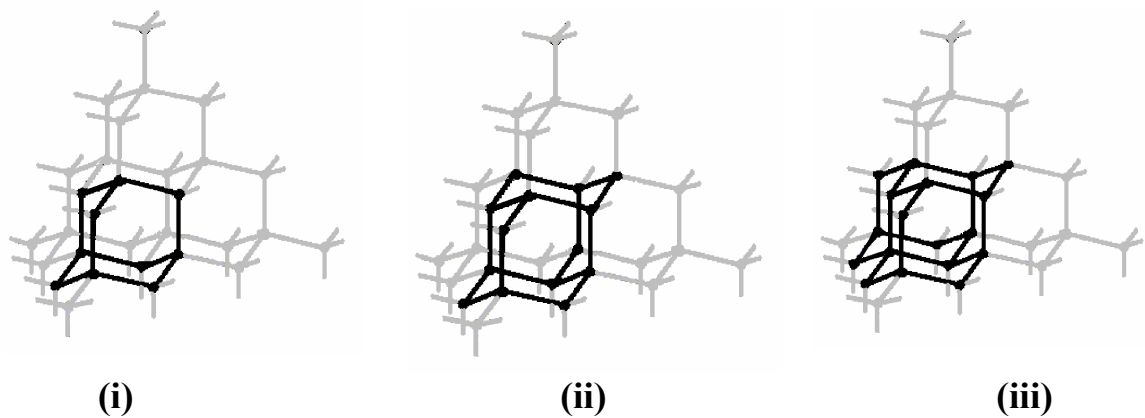

**(i)**     **(ii)**     **(iii)**

**Figure 5:** The relation between lattice diamond structure and (i). adamantane, (ii). diamantane and (iii), triamantane structures.

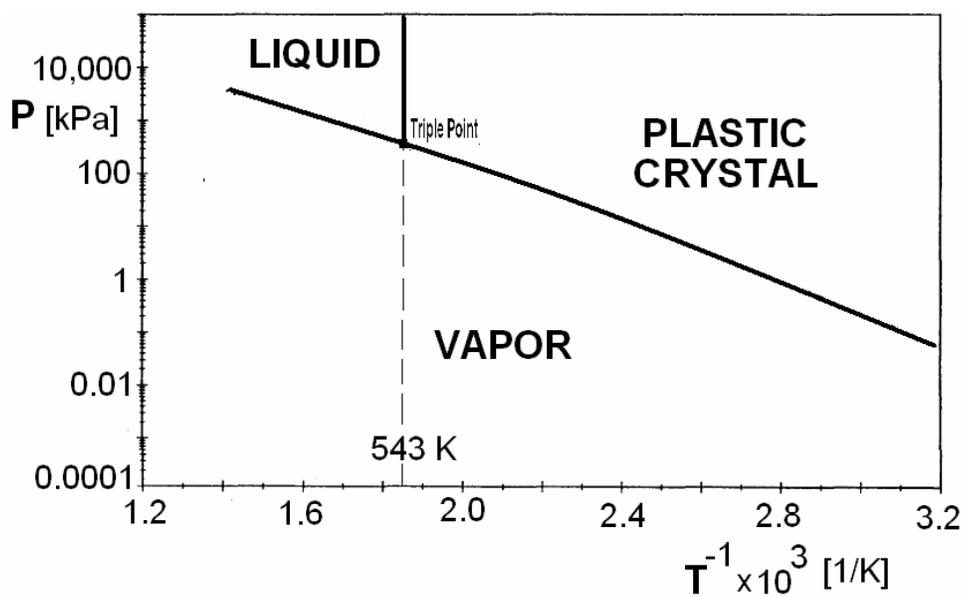

**Figure 6:** Vapor-Liquid-Solid (plastic crystal) phase diagram of adamantane. The phase transition from plastic crystal to rigid crystal phase occurs at 208.6 K (1/T=0.004794 K$^{-1}$). This diagram is based on the data of Table II.

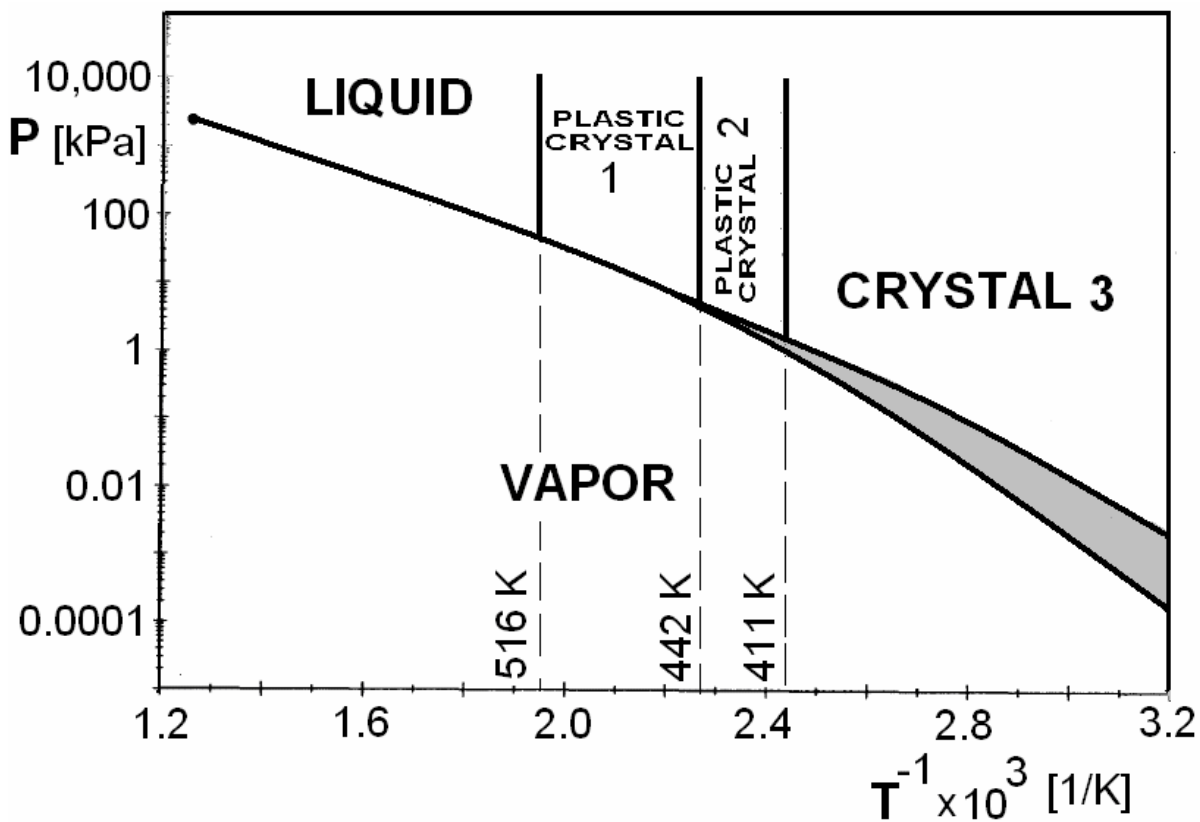

**Figure 7:** Vapor-Liquid-Solids (plastic crystal 2, plastic crystal 2, crystal 3) phase diagram of diamantane. This diagram is based on the data of Table II. The shaded area between vapor and plastic crystal 2 & crystal 3 phase transitions is indicative of the error range of the available data**.**

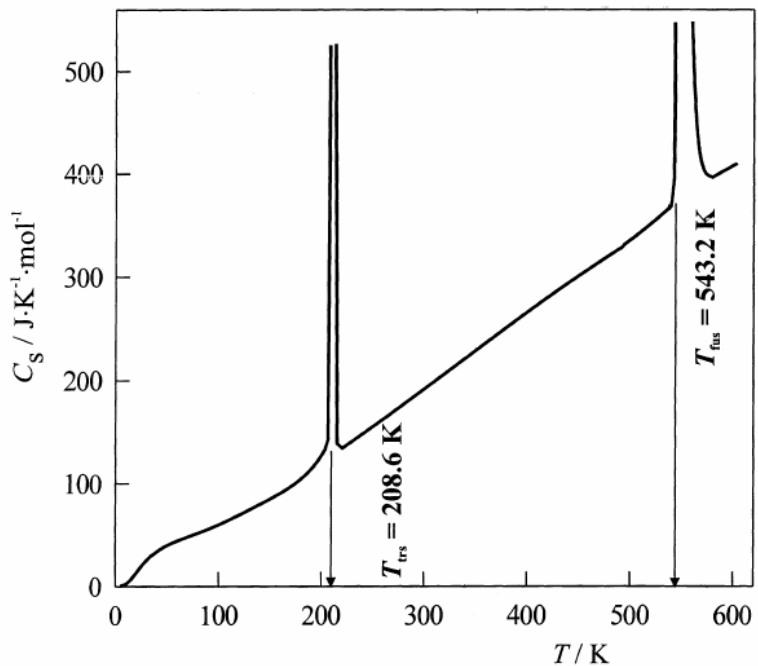

**Figure 8:** The temperature dependence of the heat capacity in the condensed state for adamantane [5] as measured by a scanning calorimeter. $T_{trs}$ stands for temperature of transition from rigid crystal (fcc)-to-plastic crystal (cubic) state of adamantane and $T_{fus}$ stands for fusion temperature.

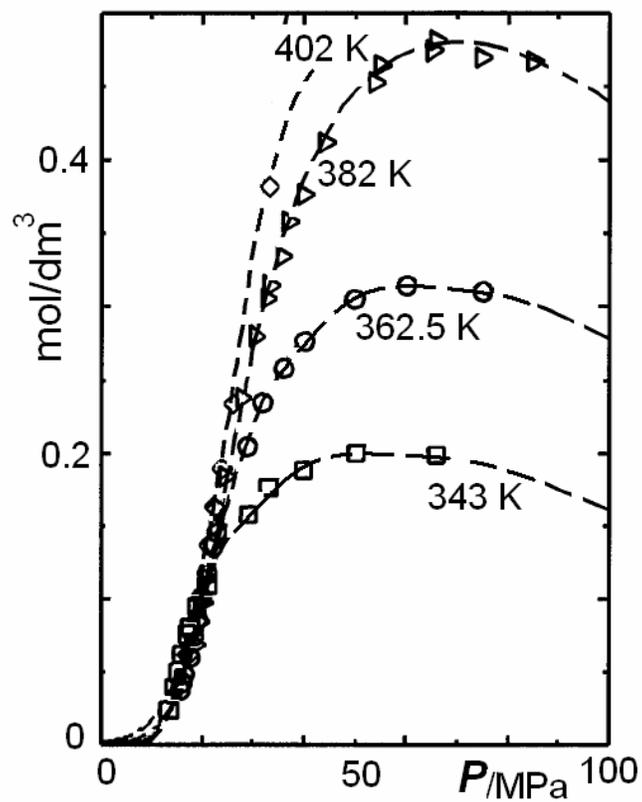

**Figure 9:** Effect of temperature and pressure on solubility (in units of mol/dm$^3$) of adamantane in dense (supercritical) carbon dioxide gas. Data from [36].

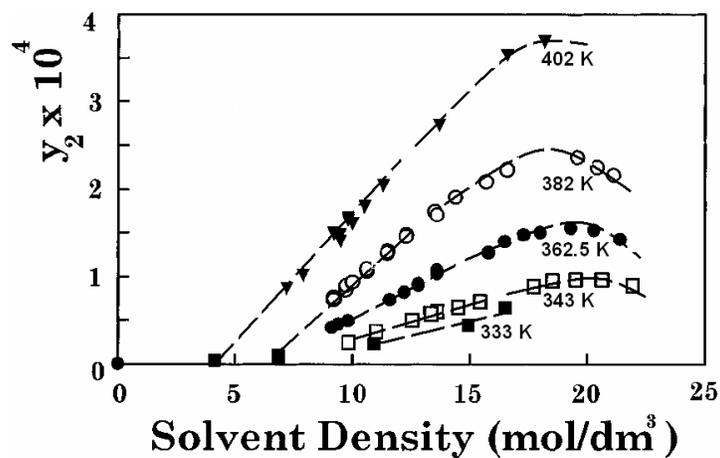

**Figure 10:** Effect of temperature and supercritical solvent density on solubility of adamantane (in units of mole fraction) in dense (supercritical) carbon dioxide. Data of isotherm at 333 K is from [35]. Data of isotherms at 343 K, 362.5 K, 382 K and 402 K are from [37].

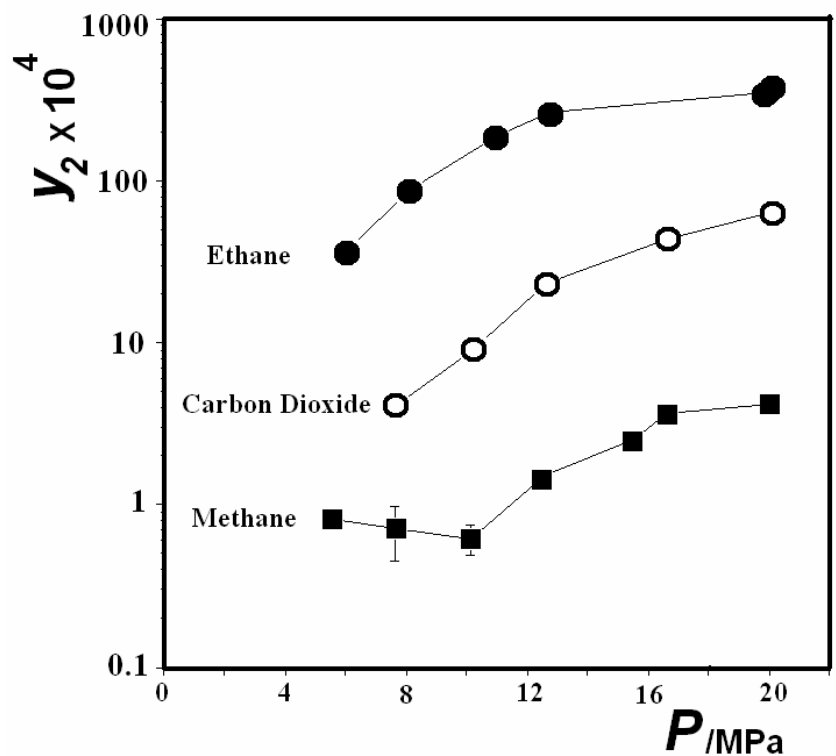

**Figure 11:** Effect of pressure on the solubility (in units of mole fraction) of adamantane in dense (supercritical) carbon dioxide, methane, and ethane gases at 333 K. Data from [35].

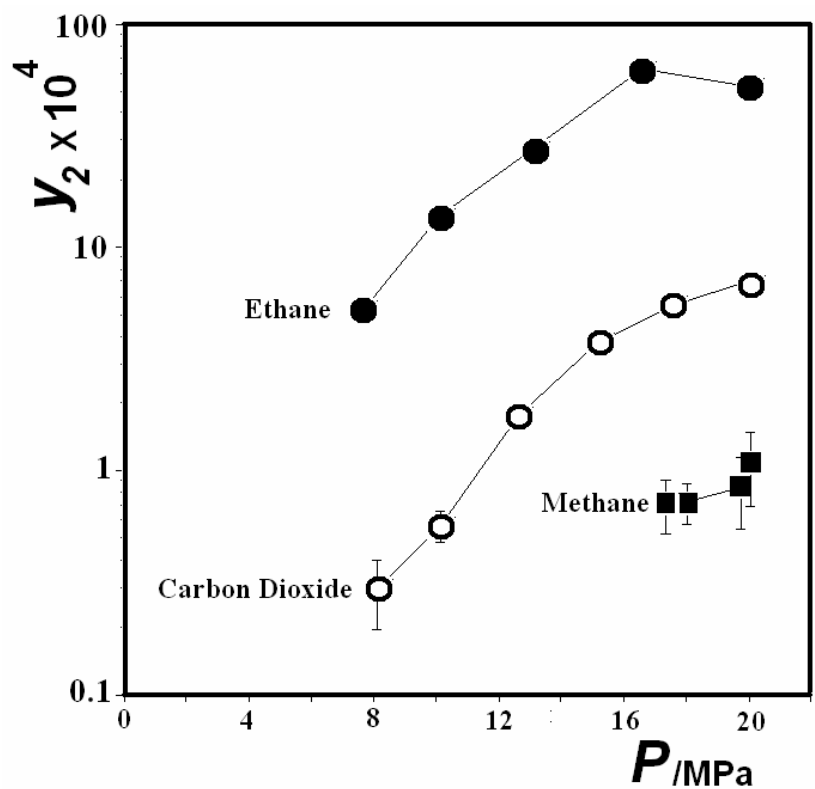

**Figure 12:** Effect of pressure on solubility (in units of mole fraction) of diamantane in dense (supercritical) gases at 333 K (for carbon dioxide and ethane) and at 353 K (for methane) – Data from [35].

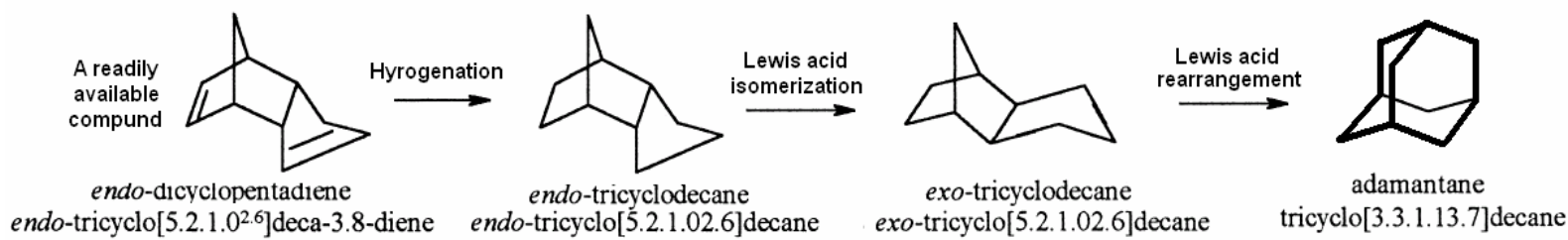

**Figure 13:** Various stages in synthesis of adamantane based on the work reported in [43,44].

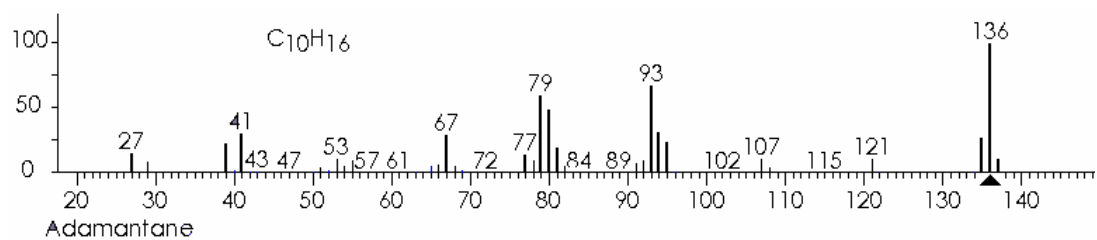

**Figure 14:** Standard molecular fragmentation spectrum of adamantane (136 m/z)

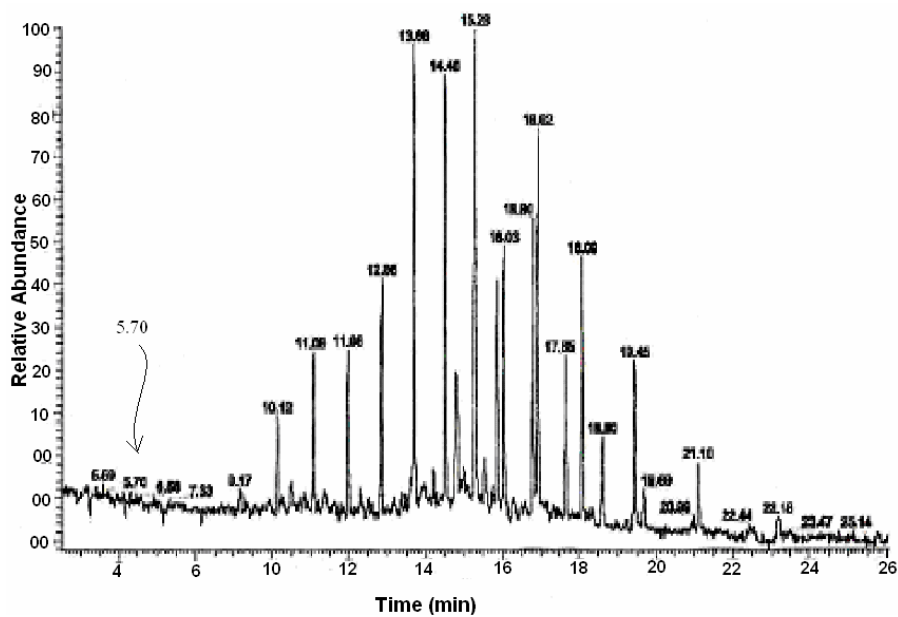

**Figure 15:** Gas chromatogram of a gas condensate (NGL=natural-gas liquid) sample [74]. The peak with retention time of 5.70 eluted between $nC_{15}$ and $nC_{16}$ is indicative of the probable existence of diamantane in the sample.

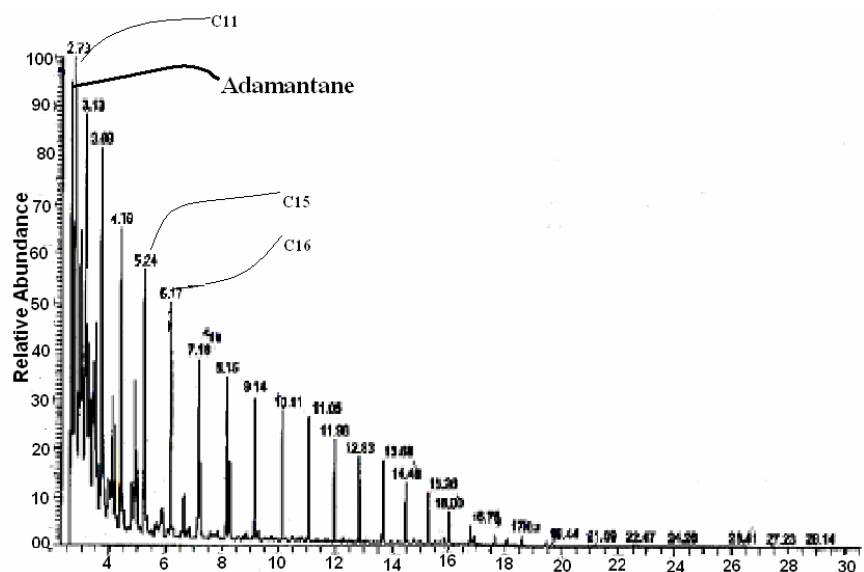

**Figure 16** – Gas chromatogram of a crude oil sample showing the possible existence of adamantane and diamantane in the sample.

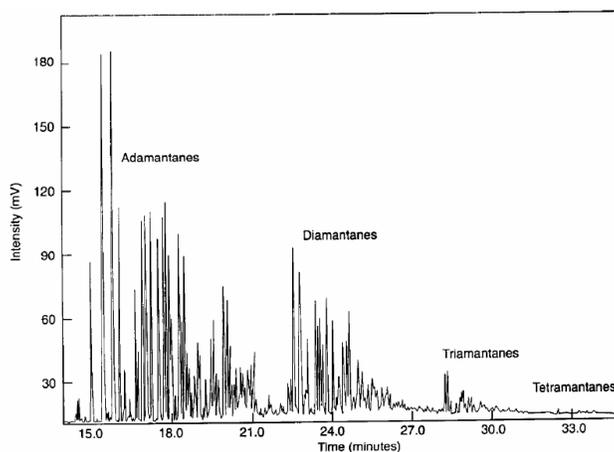

**Figure 17:** Gas chromatogram of a diamondoid-rich gas-condensate (NGL) sample showing clusters of peaks representing adamantanes, diamantanes, triamantanes and tetramantanes. From: [11].

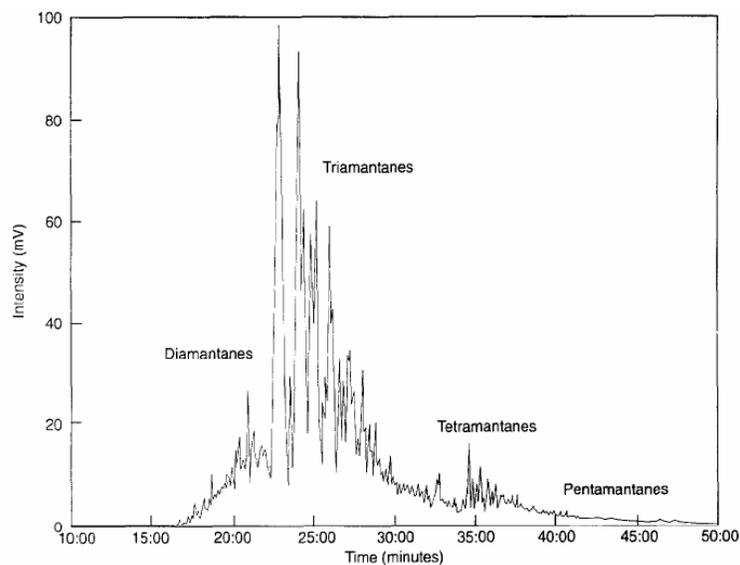

**Figure 18:** Gas chromatogram from the full-scan CC/MS analysis of a high-temperature distillation fraction (343" $C^+$) containing diamondoids. From: [11].

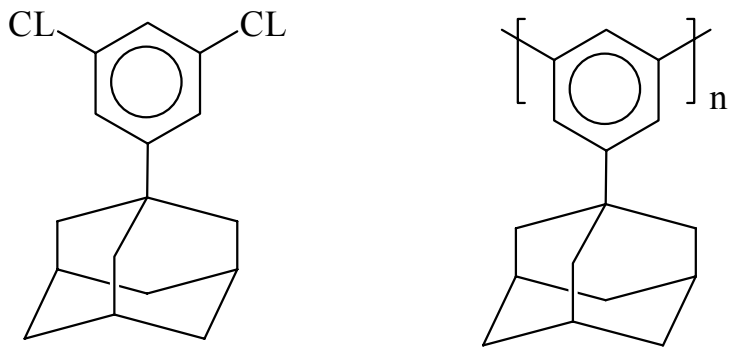

**Figure 19:** (Left) 1,3-dichloro-5-(1-admantyl) benzene monomer and (Right) adamantly-substituted poly(m-phenylene) which is shown to have a high degree of polymerization and stability decomposing at high temperatures of around 350 °C. From [89].

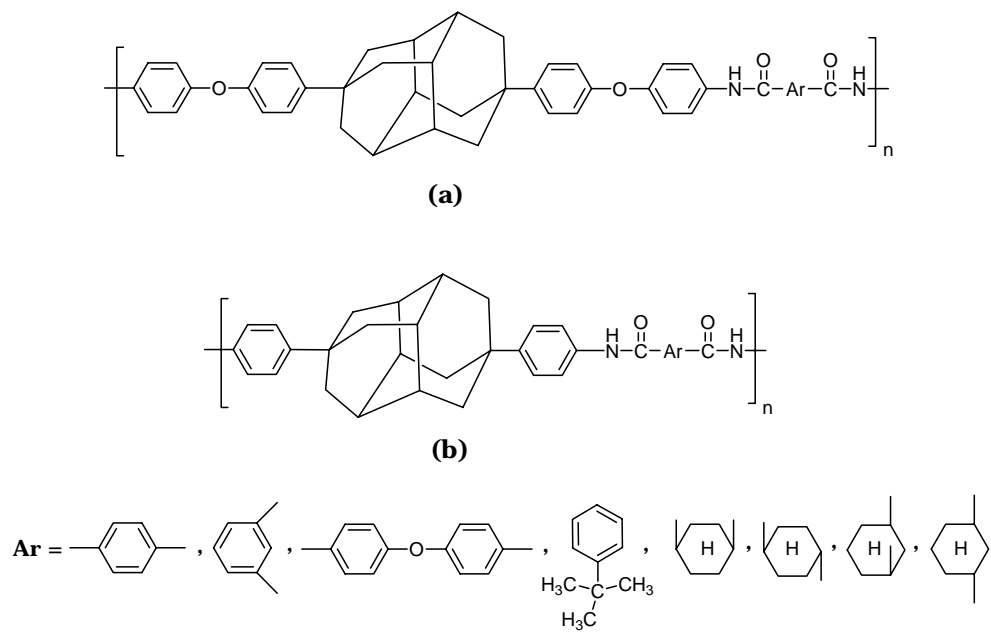

**Figure 20:** Diamantane-based polyamides; (a) derived from 4,9-bis[4-(4-aminophenoxy)phenyl]diamantane and (b) derived from 4,9-bis(4-aminophenyl)diamantane. From [90].

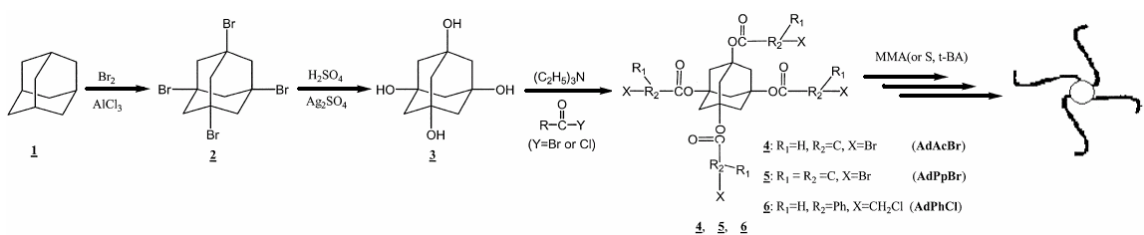

**Figure 21:** Atom transfer radical polymerization (ATRP) synthetic route to tetrafunctional initiators of a star polymer with adamantly (adamantane core). From [91].

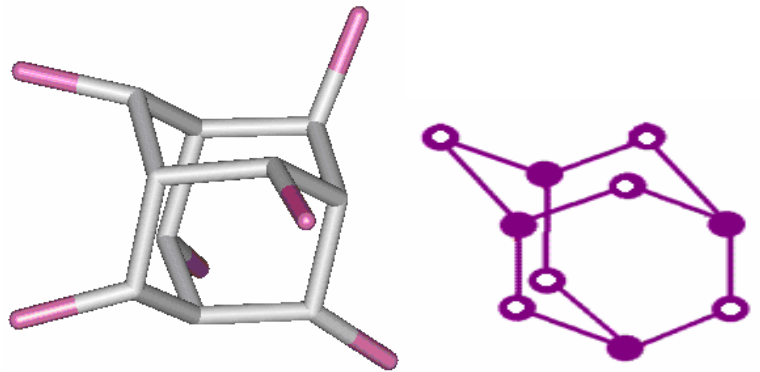

**Figure 22:** Demonstration of the six linking groups of adamantane.

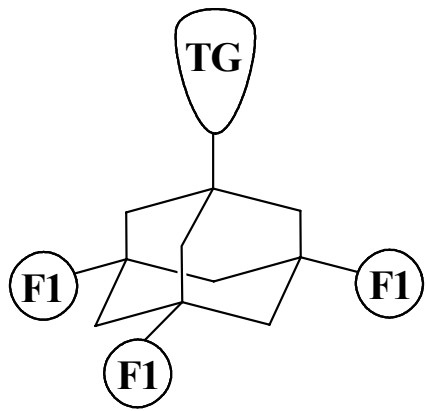

**Figure 23:** Schematic drawing which shows adamantane as a molecular probe with three fluorophore groups (F1) and a targeting part (TG) for specific molecular recognition. From [112].

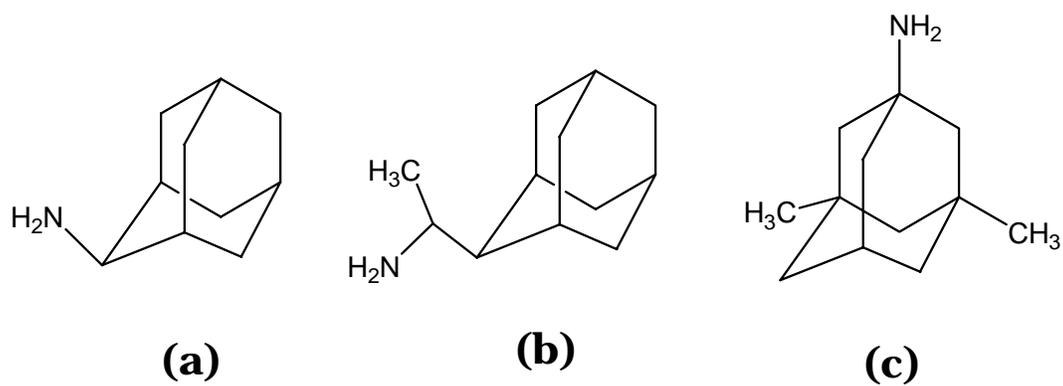

**(a)**          **(b)**          **(c)**

**Figure 24:** Chemical structures of (a) Amantadine. (b) Rimantadine. (c) Memantine.

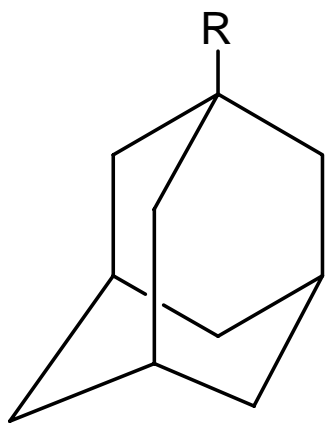

**(I) R = NH$_2$-Tyr-(D-Ala)-Gly-Phe-Leu-CO-O-**
**(II)R = NH$_2$-Tyr-(D-Ala)-Gly-Phe-Leu-CO-NH-**

**Figure 25:** The adamantane-conjugated [D-Ala2]Leu-enkephalin prodrugs. From [144].

Polymer (1)

Polymer (2)

**Figure 26:** Polymer (1) {poly (ethyladamantyl β-malate)} is hydrophobic and polymer (2) {poly(β-malic acid-co-ethyladamantyl β-malate} is hydrophilic. Both of these plymers are used as carriers for different drugs [148].

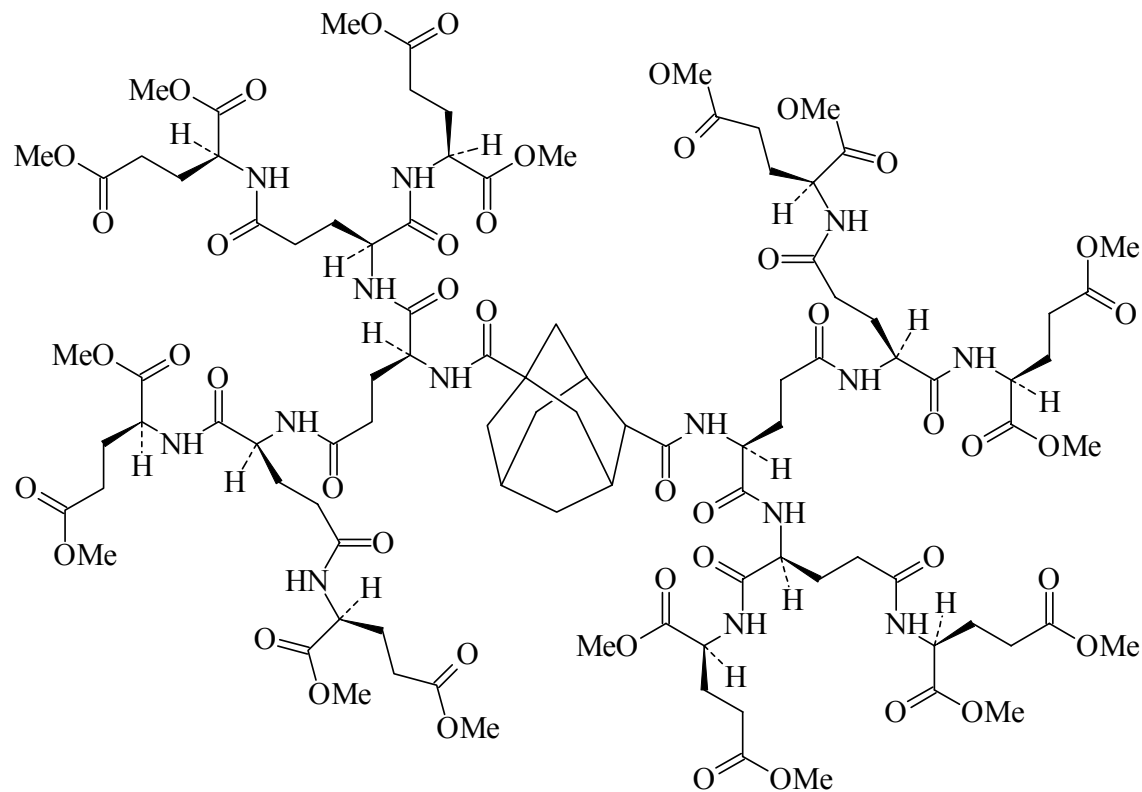

**Figure 27:** Adamantane nucleus with amino acid substituents creates a peptidic matrix [151]. The represented structure is Glu4-Glu2-Glu-[ADM]-Glu-Glu2-Glu4.

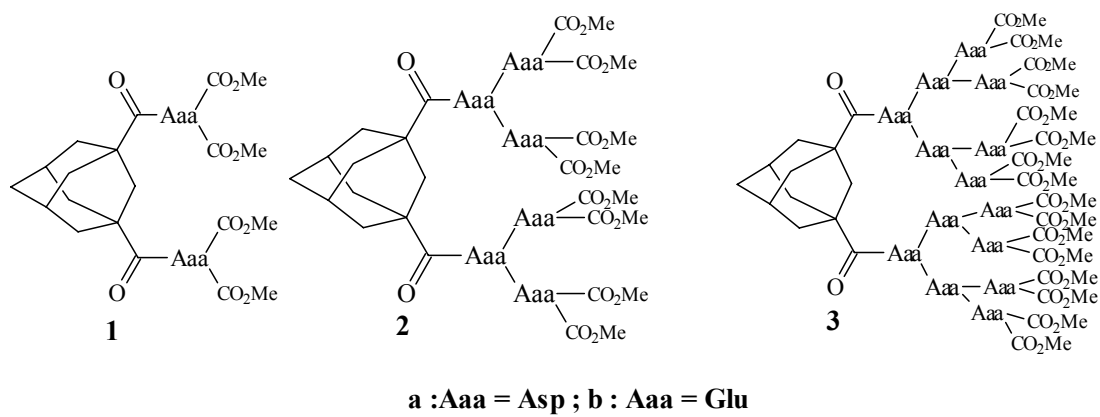

**a :Aaa = Asp ; b : Aaa = Glu**

**Figure 28:** A dendrimer-based approach for the design of globular protein mimic using glutamic (Glu) and aspartic (Asp) acids as the building blocks and adamantyl as the core [151].

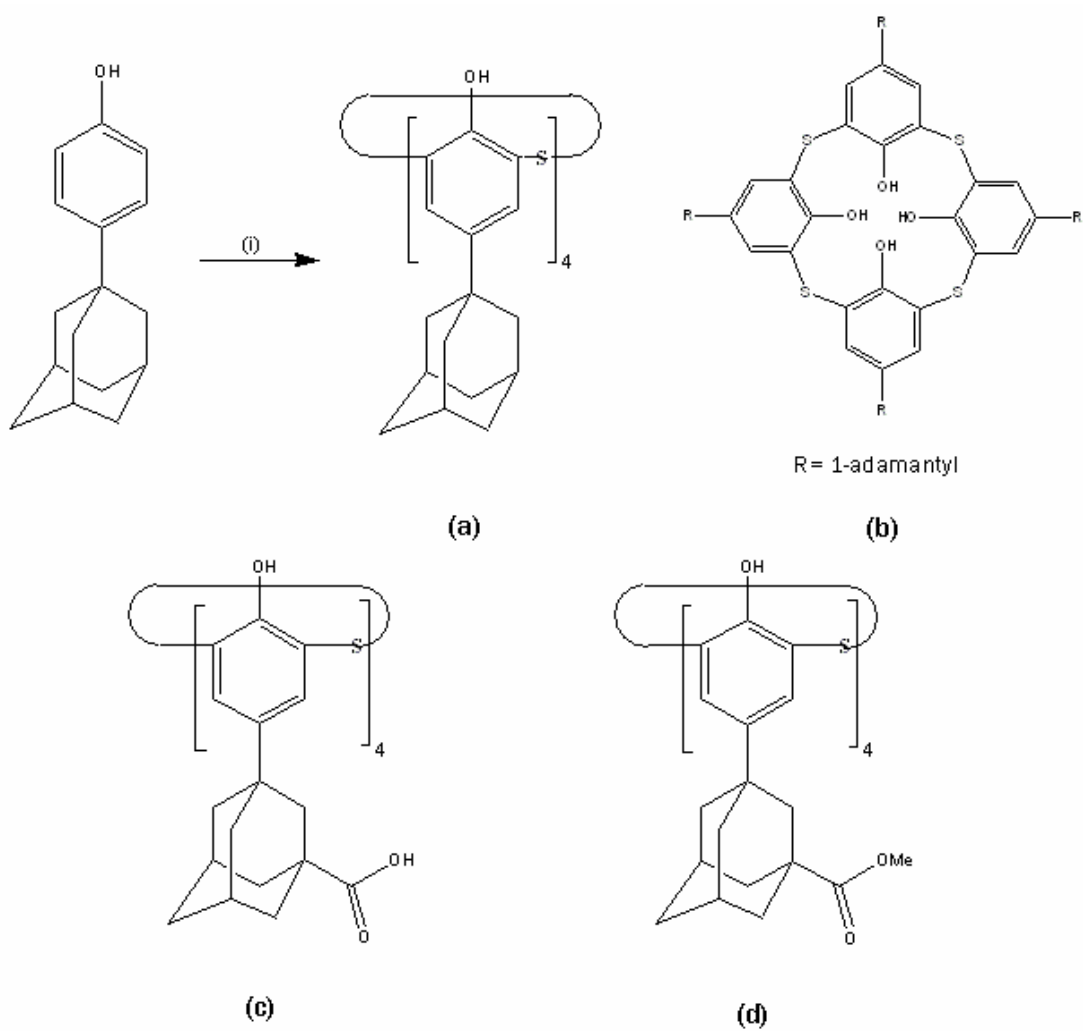

**Figure 29:** a) Synthesis route of the molecule (b): (i) S8, NaOH, tetraethyleneglycol dimethyl ether, heat, (28%). b) Adamantane Upper rim derivative based on the thiacalix[4]arene platform. c,d) The carboxylic acid and ester derivative of adamantane can be also used as substituents. From [109].

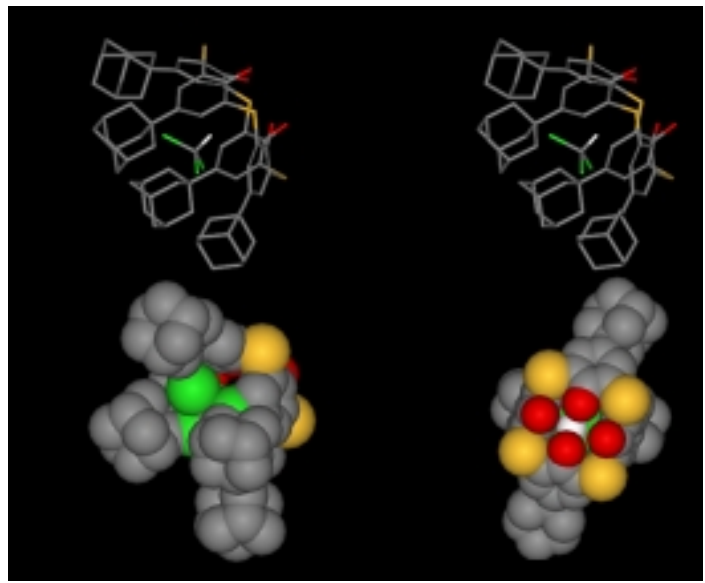

**Figure 30:** Lateral stereo views of adamantane-derivative thiacalix[4]arene (top) presented in Figure 29. A $CHCl_3$ molecule has been entrapped inside the inclusion compound. The bottom view (left bottom) and top view (right bottom) have been also shown. H atoms have been removed from the inclusion compound for more clarity. (Cl, OH, S, H and C atoms have been colored green, red, yellow, white and gray, respectively).

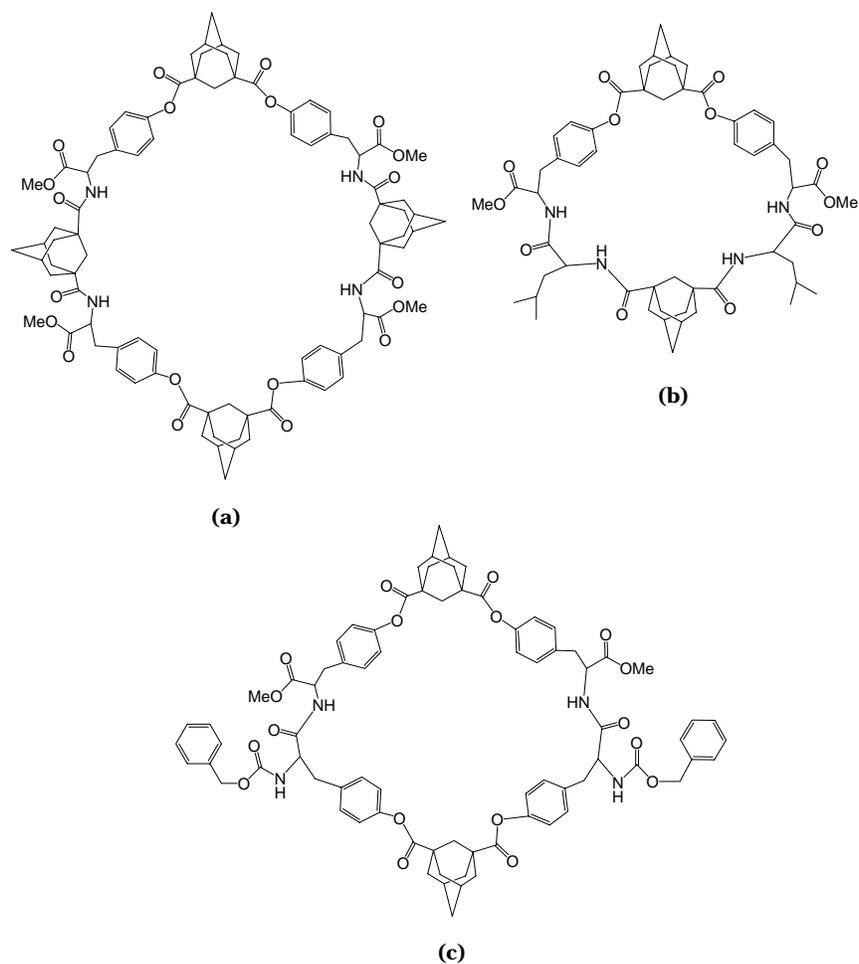

**Figure 31:** Adamantane-bridged tyrosine-based cyclodepsipeptides are suitable models for host-guest studies and they are also able to act as ion transporters. From [161].

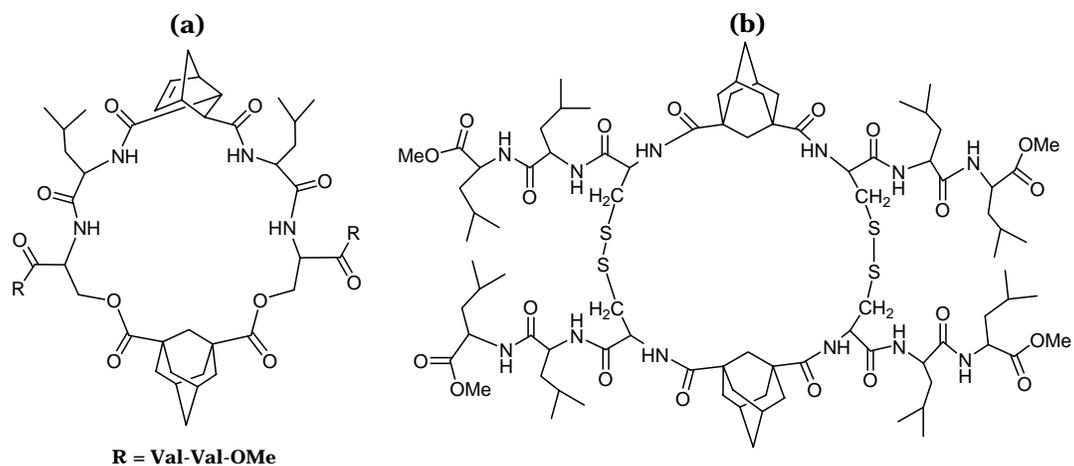

**R = Val-Val-OMe**

**Figure 32:** Adamantane-containing norbornene ( 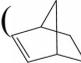 )-constrained cyclic peptides possess the ability to transport ions across the model membranes in both specific and non-specific ways [162].

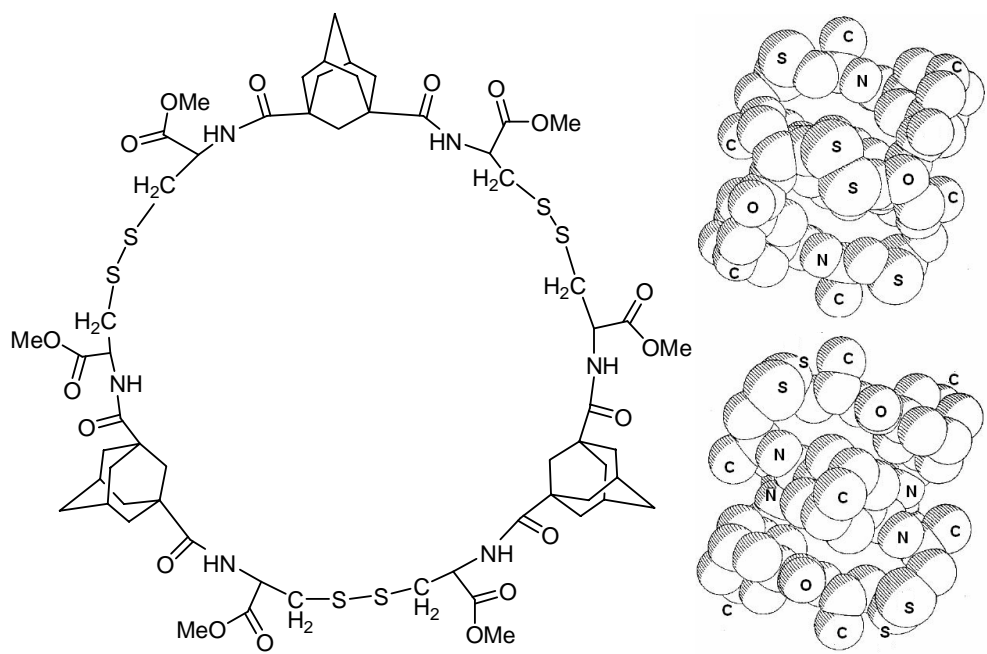

**Figure 33:** The cyclo (Adm-Cyst)3 as adopts a figure-eight like helical structure. The chiral amino acid, cystine, configuration determines the helix disposition (right-handed or left handed helix). Adamantane plays an important role as a ring size controlling agent. From [163].

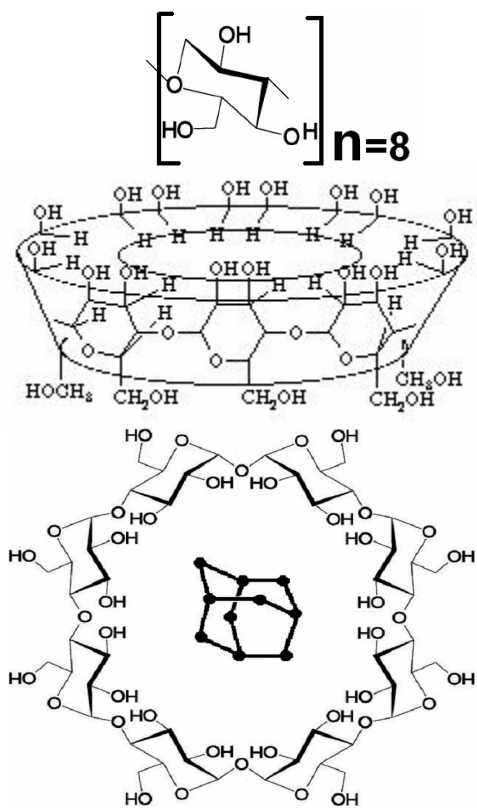

**Figure 34:** Chemical formula of γ cyclodextrin consisting of eight glucose molecules with adamantane as is the guest entrapped within its hydrophobic cavity. Structures of α and β cyclodextrins will be similar but made up of six and seven (n=6, 7) glucoses, respectively

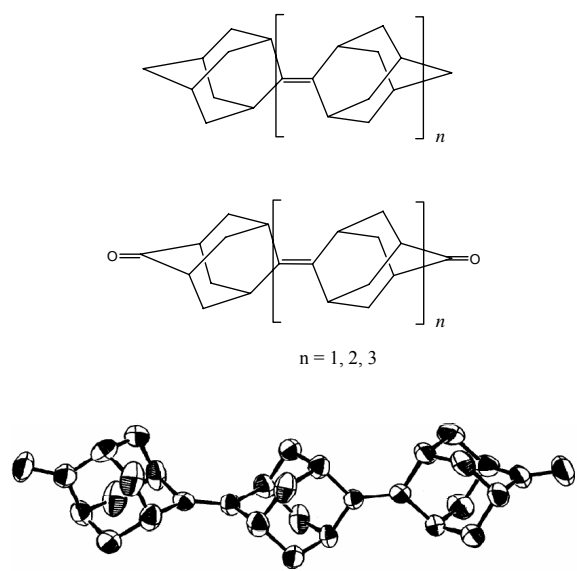

**Figure 35:** Poly-adamantane molecular rods. From [170].

_______________________

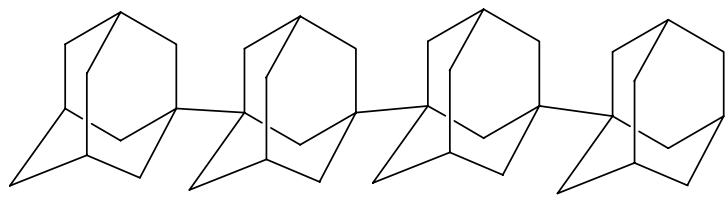

**Figure 36:** Synthetic design of a molecular rod made of adamantanes: The tetrameric 1,3-adamantane. From [171].